\documentclass[12pt,preprint]{aastex}
\usepackage{hyperref}

\shorttitle{Reconstructing CMEs with Imaging and In Situ Data}

\shortauthors{Liu et al.}

\begin{document}

\title{Reconstructing CMEs with Coordinated Imaging and In Situ
Observations: Global Structure, Kinematics, and Implications for
Space Weather Forecasting}

\author{Ying Liu\altaffilmark{1}, Arnaud Thernisien\altaffilmark{2},
Janet G. Luhmann\altaffilmark{1}, Angelos Vourlidas\altaffilmark{3},
Jackie A. Davies\altaffilmark{4}, Robert P. Lin\altaffilmark{1,5},
and Stuart D. Bale\altaffilmark{1}}

\altaffiltext{1}{Space Sciences Laboratory, University of
California, Berkeley, CA 94720, USA; liuxying@ssl.berkeley.edu}

\altaffiltext{2}{Universities of Space Research Association,
Columbia, MD 21044, USA}

\altaffiltext{3}{Space Science Division, Naval Research Laboratory,
Washington, DC 20375, USA}

\altaffiltext{4}{Space Science and Technology Department, Rutherford
Appleton Laboratory, Didcot, UK}

\altaffiltext{5}{School of Space Research, Kyung Hee University,
Yongin, Gyeonggi 446-701, Korea}

\begin{abstract}

We reconstruct the global structure and kinematics of coronal mass
ejections (CMEs) using coordinated imaging and in situ observations
from multiple vantage points. A forward modeling technique, which
assumes a rope-like morphology for CMEs, is used to determine the
global structure (including orientation and propagation direction)
from coronagraph observations. We reconstruct the corresponding
structure from in situ measurements at 1 AU with the Grad-Shafranov
(GS) method, which gives the flux-rope orientation, cross section
and a rough knowledge of the propagation direction. CME kinematics
(propagation direction and radial distance) during the transit from
the Sun to 1 AU are studied with a geometric triangulation
technique, which provides an unambiguous association between solar
observations and in situ signatures; a track fitting approach is
invoked when data are available from only one spacecraft. We show
how the results obtained from imaging and in situ data can be
compared by applying these methods to the 2007 November 14-16 and
2008 December 12 CMEs. This merged imaging and in situ study shows
important consequences and implications for CME research as well as
space weather forecasting: (1) CME propagation directions can be
determined to a relatively good precision as shown by the
consistency between different methods; (2) the geometric
triangulation technique shows a promising capability to link solar
observations with corresponding in situ signatures at 1 AU and to
predict CME arrival at the Earth; (3) the flux rope within CMEs,
which has the most hazardous southward magnetic field, cannot be
imaged at large distances due to expansion; (4) the flux-rope
orientation derived from in situ measurements at 1 AU may have a
large deviation from that determined by coronagraph image modeling;
(5) we find, for the first time, that CMEs undergo a westward
migration with respect to the Sun-Earth line at their acceleration
phase, which we suggest as a universal feature produced by the
magnetic field connecting the Sun and ejecta. Importance of having
dedicated spacecraft at L4 and L5, which are well situated for the
triangulation concept, is also discussed based on the results.

\end{abstract}


\section{Introduction}

Coronal mass ejections (CMEs) are the most spectacular eruptions in
the solar corona in which $10^{15-16}$ g of plasma with $10^{31-32}$
ergs of energy is hurled into interplanetary space
\citep[e.g.,][]{gosling74, hundhausen97}. The ejected materials in
the solar wind, a key link between activity at the Sun and
disturbances in the heliosphere, are called interplanetary coronal
mass ejections (ICMEs). A subset of ICMEs, termed as magnetic clouds
(MCs), are characterized by a strong magnetic field, a smooth and
coherent rotation of the field, and a depressed proton temperature
compared with the ambient solar wind \citep{burlaga81}.

CMEs have been recognized as drivers of major space weather effects.
They are responsible for the most intense solar energetic particle
events, which can endanger life and technology on the Earth and in
space. They can also cause major geomagnetic storms in the
terrestrial environment including large auroral currents and high
particle fluxes, which can disrupt satellite operations, power
systems and radio communications. CMEs drive space weather effects
typically in two ways. First, CMEs are often associated with a
sustained southward magnetic field, which can reconnect with
geomagnetic fields and produce storms in the terrestrial environment
\citep{dungey61, gosling91}. The southward field component is either
within the ejecta or produced by the interaction of the ejecta with
the ambient medium \citep{gosling87, mccomas88, liu08a}. Second,
fast CMEs can generate interplanetary shocks, a key source of
energetic particles and radio bursts. Note that the magnetic
reconnection rate at the dayside of the magnetopause is controlled
by the dawn-dusk electric field, $-\textbf{v}_r\times \textbf{B}_s$,
where $\textbf{v}_r$ is the radial velocity of the solar wind and
$\textbf{B}_s$ is the southward magnetic field component. Therefore,
the high speed of CMEs can also significantly enhance the magnetic
reconnection rate when a southward magnetic field is present.

CMEs have been studied by remote sensing of these events at the Sun
and by in situ measurements of their plasma properties when they
encounter spacecraft. However, most CME studies are focused on the
phenomena either at the Sun or near the Earth; efforts to link solar
and in situ observations, especially the development of practical
strategies for space weather forecasting, are still lacking. The
global structure of CMEs and the underlying physics governing CME
propagation in the heliosphere are not well understood. Remote
sensing observations, mostly by coronagraphs, can provide a long
warning time in terms of the occurrence and speed of CMEs.
Coronagraphs record photospheric radiation (or white light)
Thomson-scattered by electrons \citep{billings66}, so what is
observed is essentially a density structure projected onto the sky.
The magnetic field orientation is difficult to determine directly
from white-light observations. It is also a challenge to infer the
three-dimensional (3D) structure and kinematics from a single
viewpoint due to projection effects. In situ measurements at the
first Lagrangian point (L1) can give accurate information about the
plasma and magnetic field structure of CMEs but only along a
one-dimensional (1D) cut through the large-scale 3D structure. A
forecasting time provided by in situ measurements at L1 is only
about 30 minutes depending on the speed of the solar wind. Long-term
and precise space weather forecasting, as well as the determination
of CME global structure and kinematics, requires coordinated imaging
and in situ observations from multiple vantage points.

Now we have several spacecraft looking at the Sun including the
Solar and Heliospheric Observatory \citep[SOHO;][]{domingo95} and
the Solar Terrestrial Relations Observatory
\citep[STEREO;][]{kaiser08}. STEREO comprises two spacecraft with
one preceding the Earth (STEREO A) and the other trailing behind
(STEREO B). Improved determination of CME global structure and
kinematics is feasible with these multiple viewpoints. In
particular, STEREO has two wide-angle heliospheric imagers (HI1 and
HI2) which can cover the whole Sun-Earth space (except the
heliospheric polar regions). Evolving CME properties determined from
imaging observations can thus be compared with in situ measurements
for a better understanding of the CME-ICME relationships. In this
work, we combine image observations with in situ measurements to
constrain the global structure and kinematics of CMEs. Implications
for CME research and space weather forecasting are discussed in
terms of consistency and caveats in linking imaging and in situ data
and future observational concept. Observations and methodology are
described in \S 2. We present details on case studies in \S 3. The
results are summarized and discussed in \S 4. We also provide two
appendices for error analysis and the discussion of various
techniques with which to convert elongation to distance,
respectively.

\section{Observations and Methodology}

This study requires joint white-light and in situ observations. We
use white-light observations from STEREO and SOHO, and in situ
plasma and magnetic field measurements from STEREO, ACE and WIND.
Each of the STEREO spacecraft carries an identical imaging suite,
the Sun Earth Connection Coronal and Heliospheric Investigation
\citep[SECCHI;][]{howardra08}, which consists of an EUV imager
(EUVI), two coronagraphs (COR1 and COR2) and two heliospheric
imagers (HI1 and HI2). COR1 and COR2 have a field of view (FOV) of
0.4$^{\circ}$ - 1$^{\circ}$ and 0.7$^{\circ}$ - 4$^{\circ}$ around
the Sun, respectively. HI1 has a 20$^{\circ}$ square FOV centered at
14$^{\circ}$ elongation from the center of the Sun while HI2 has a
70$^{\circ}$ FOV centered at 53.7$^{\circ}$. Combined together these
cameras can image a CME from its birth in the corona all the way to
the Earth and beyond \citep[see][and the animations online]{liu09a,
liu10}. STEREO also has several sets of in situ instrumentation,
including the In Situ Measurements of Particles and CME Transients
(IMPACT) package \citep{luhmann08} and the Plasma and Suprathermal
Ion Composition (PLASTIC) investigation \citep{galvin08}, which
provide in situ measurements of the magnetic field, particles and
the bulk solar wind plasma. At L1, the Large Angle Spectroscopic
Coronagraph \citep[LASCO;][]{brueckner95} aboard SOHO gives another
view of the Sun and ACE and WIND monitor the near-Earth solar wind
conditions, thus adding a third vantage point.

\subsection{Image Forward Modeling}

CME coronagraph images can be well reproduced by a forward modeling
technique with a geometric model \citep{thernisien06, thernisien09}.
The model adopts a rope-like morphology for CMEs with two ends
anchored at the Sun. A number of free parameters are used in the
model to control the global shape of the rope. An ad hoc electron
density distribution is generated through the rope, and then
synthetic images are derived from the density distribution using a
ray-tracing program. Comparison between modeled images and observed
ones from different vantage points can give the rope orientation and
propagation direction. To save computing time, the overall shape
with a wireframe rendering can be compared with observations without
calculating the brightness. The model assumes a self-similar
expansion for the rope, so propagation of CMEs can be simulated by
just varying the height of the rope. This forward modeling technique
is less useful for CMEs when the signal becomes too weak for a
reliable delineation of the outer CME envelope and/or distortion of
CMEs by the ambient structures becomes significant.

A good visual agreement with observed images can be obtained by
adjusting those parameters. Note that, even though the model has a
flux-rope geometry, essentially it is a density structure. We will
relate the forward modeling with in situ reconstruction to see if
information about the magnetic field orientation can be inferred
from imaging observations.

\subsection{In Situ Reconstruction}

Correspondingly, we can attempt to reconstruct the structure using
in situ data if a CME encounters spacecraft. Initially designed for
the study of the terrestrial magnetopause \citep[e.g.,][]{hau99},
the Grad-Shafranov (GS) technique can be applied to flux-rope
reconstruction \citep[e.g.,][]{hu02}. The key idea of this method is
that the thermal pressure and axial magnetic field depend on the
vector magnetic potential only, which has been validated by well
separated multi-spacecraft measurements \citep{liu08c}. The
advantage of the GS technique is that it relaxes the force-free
assumption and can give a cross section as well as flux-rope
orientation without prescribing the geometry. Velocity and magnetic
field measurements within an MC are transformed into a
deHoffmann-Teller (HT) frame in which the electric field vanishes
\citep[e.g.,][]{khrabrov98}. MHD equilibrium obtained in this frame
results in the GS equation under the assumption of a translational
symmetry \citep[e.g.,][]{schindler73, sturrock94}. The axis
orientation of an MC is determined from the single-valued behavior
of the thermal pressure and axial magnetic field over the vector
potential \citep{hu02}. Once the axis orientation is acquired, a
flux-rope frame is set up with the $z$ direction along the flux-rope
axis. The GS equation is then solved for the vector potential in
this flux-rope frame using in situ measurements as boundary
conditions, which yields a cross section in a rectangular domain.

We will compare the in situ reconstruction with the forward modeling
of CME coronagraph images. The goal is to examine how the comparison
between in situ reconstruction and image modeling constrains the
global structure, propagation direction and orientation of CMEs.
Effects such as solar wind distortion and flux-rope rotation between
the Sun and 1 AU may also be addressed. The relationship between CME
and ICME geometries may lead to a possible prediction of the
magnetic field orientation within the ejecta at 1 AU from solar
observations, which is important for space weather forecasting.

\subsection{CME Tracking}

Association between solar observations and in situ signatures at 1
AU is often ambiguous due to the large distance gap. CMEs may also
change appreciably as they propagate in interplanetary space. To
link the solar and in situ observations, essentially we need to
track CMEs continuously over a large distance. Here we use a
geometric triangulation technique, which can determine both
propagation direction and radial distance of CMEs with stereoscopic
imaging observations from STEREO \citep[][hereinafter referred to as
paper 1]{liu10}. In paper 1, we focused on CME propagation between
STEREO A and B, but the same concept can also be applied to CMEs
propagating outside the space between the two spacecraft (currently
limited to coronagraphs). Figure~1 shows the configuration in the
ecliptic plane used for the geometric triangulation. STEREO A is
slightly closer to the Sun than the Earth and leads the Earth, while
STEREO B is a little further and trails the Earth. Each spacecraft
drifts away from the Earth at a rate of about 22.5$^{\circ}$ per
year. These two spacecraft make independent measurements of the
elongation angle of a CME feature (the angle of the feature with
respect to the Sun-spacecraft line), denoted as $\alpha_A$ and
$\alpha_B$ for STEREO A and B respectively. The behavior of the
elongation angle as viewed from different vantage points forms the
basis to determine the propagation direction and radial distance.

The simple geometry shown in Figure~1 gives
\begin{equation}
\frac{r\sin(\alpha_A+\beta_A)}{\sin\alpha_A} = d_A,
\end{equation}
\begin{equation}
\frac{r\sin(\alpha_B+\beta_B)}{\sin\alpha_B} = d_B,
\end{equation}
where $r$ is the radial distance of the feature from the Sun,
$\beta_A$ and $\beta_B$ are the propagation angles of the feature
relative to the Sun-spacecraft line, and $d_A$ and $d_B$ are the
distances of the two spacecraft (known). Note that $\alpha$ and
$\beta$ are all positive and $\alpha < \pi/2$. We also have
\begin{mathletters}
\begin{equation}
\beta_A + \beta_B = \gamma,
\end{equation}
\begin{equation}
\beta_A - \beta_B = \gamma,
\end{equation}
\begin{equation}
\beta_B - \beta_A = \gamma,
\end{equation}
\end{mathletters}
for a feature propagating between (Figure~1, left), east of
(Figure~1, middle), and west of (Figure~1, right) the two
spacecraft, respectively. Here $\gamma$ is the longitudinal
separation of the two spacecraft (also known). For CMEs that are
directed away from the Earth (i.e., backsided), the above equation
becomes $\beta_A + \beta_B = 2\pi - \gamma$, but those events would
be of no interest from the perspective of space weather prediction.
These equations can be reduced to
\begin{mathletters}
\begin{equation}
\tan\beta_A = \frac{\sin\alpha_A\sin(\alpha_B+\gamma) -
f\sin\alpha_A\sin\alpha_B}{\sin\alpha_A\cos(\alpha_B+\gamma) +
f\cos\alpha_A\sin\alpha_B},
\end{equation}
\begin{equation}
\tan\beta_A = \frac{\sin\alpha_A\sin(\alpha_B-\gamma) -
f\sin\alpha_A\sin\alpha_B}{-\sin\alpha_A\cos(\alpha_B-\gamma) +
f\cos\alpha_A\sin\alpha_B},
\end{equation}
\begin{equation}
\tan\beta_A = \frac{\sin\alpha_A\sin(\alpha_B+\gamma) -
f\sin\alpha_A\sin\alpha_B}{-\sin\alpha_A\cos(\alpha_B+\gamma) +
f\cos\alpha_A\sin\alpha_B},
\end{equation}
\end{mathletters}
for the three cases separately, where $f = d_B/d_A$ (which varies
between 1.04 and 1.13 during a full orbit of the STEREO spacecraft
around the Sun). This equation allows a quick estimate of the
propagation direction.

A first step would be to determine which equation (4a, 4b, or 4c)
should be used; whether a CME is propagating between, east or west
of the two spacecraft can be easily identified from coronagraph
observations. The elongation angles can be obtained from
time-elongation maps produced by stacking the running difference
intensities along the ecliptic plane (see paper 1 and details
below). Even weak signals are discernible in these maps, so
transient activity can be revealed over an extensive region of the
heliosphere. CME features (e.g., leading or trailing edges with an
enhanced density) usually appear as tracks extending to large
elongation angles in the maps. Once the elongation angles
($\alpha_A$ and $\alpha_B$) are measured from the tracks, the
appropriate set of equations can then be solved for $r$, $\beta_A$
and $\beta_B$, a unique solution. The advantage of the method is
that, first, it can be applied to weak features at large distances
when time-elongation maps are used; second, it has no free
parameters (which would bring about uncertainties in the solution);
third, it can determine the propagation direction and radial
distance of CME features continuously from the Sun to a large
distance. It is also clear that the method does not require a lot of
data and the calculation is very simple. Even with a single image
pair from the two spacecraft, it can quickly estimate the
propagation direction and distance as long as the elongation angles
can be measured from the image pair.

At large distances the structures seen by the two spacecraft may
begin to bifurcate (i.e., not exactly the same part of the CME).
This situation could be worse for very wide CMEs. See Appendix B and
paper 1 for a detailed discussion of the effect of CME geometry on
the triangulation analysis. The triangulation technique, however,
still shows a reasonable accuracy in the determination of CME
kinematics (see paper 1 and details below). The combined effects due
to projection, Thomson scattering and CME geometry are expected to
be minimized for Earth-directed events (i.e., propagating
symmetrically relative to the two spacecraft). This seems to be
confirmed by our preliminary statistical study with joint imaging
and in situ data (see
\url{http://sprg.ssl.berkeley.edu/~liuxying/CME_catalog.htm}). A
practical and real-time space weather forecasting requires a means
which should be simple, efficient and easy to use. This method
satisfies all these needs. Appendix B describes another
triangulation notion under a harmonic mean approximation, which
takes into account the effect of CME geometry but at the price of
incurring significant assumptions and complications (see details in
Appendix B).

Geometric triangulation can usually be applied to coronagraph data
from the two spacecraft, but may not be possible for the HIs since
their FOVs are off to the sides of the Sun. When data are available
from only one spacecraft, we will use a track fitting approach to
determine the propagation direction and distance
\citep[e.g.,][]{sheeley99, sheeley08}. Equation (1) or (2) can be
reduced to
\begin{equation}
\alpha = \arctan\left(\frac{r\sin\beta}{d-r\cos\beta}\right).
\end{equation}
This is the so-called fixed $\beta$ \citep[or fixed $\phi$ using the
terminology of][]{kahler07} approximation. See Appendix B for
details about this approximation. A kinematic model with free
parameters is needed to fit the tracks in the time-elongation maps.
To reduce the number of free parameters, we assume that CMEs
propagate at a constant speed along a fixed radial direction in the
FOV of the HIs. This is likely true if the interaction between the
ejecta and the ambient medium is not significant. The propagation
direction ($\beta$) and radial speed can then be estimated from the
track fitting. Note that we apply this track fitting approach only
to HI data.

\section{Case Studies}

We apply the above techniques to several events to demonstrate how
the global structure and kinematics of CMEs can be constrained by
joint imaging and in situ data. Solar observations will be connected
to in situ signatures by tracking CME propagation in interplanetary
space, which provides an unambiguous association between CMEs and
ICMEs. The results from coronagraph image modeling can then be
compared with in situ reconstruction once the correspondence has
been established. To reconstruct the in situ structure, we need to
look at events that have organized magnetic fields at 1 AU (i.e.,
MCs). The 2007 November and 2008 December events are well situated
for this investigation.

\subsection{2007 November 14-20 Events}

Three consecutive CMEs are observed on 2007 November 14-16 when
STEREO A and B are separated by $\gamma\simeq40^{\circ}$. During
this time, $d_A\simeq 0.97$ AU, and $d_B\simeq 1.04$ AU. Figure~2
shows two synoptic views of the events from the two spacecraft. The
first and third events, which occur on November 14 and 16
respectively, appear at the west limb of the Sun for both STEREO A
and B, so the scenario shown in the right panel of Figure~1 will
apply. Apparently, these two CMEs show different elongation angles
for the two spacecraft, which can be converted to propagation
direction and radial distance using the geometric triangulation
method. Neither of these two events is observed by the HI
instruments on STEREO A since their FOVs are off to the east. The
second CME, which occurs on November 15, is observed by both STEREO
A and B but at opposite sides of the Sun, so it is likely an
Earth-directed event. For STEREO B, the first and second CMEs can
even be seen in HI2 whereas the third one is only faintly visible in
HI1 (see Figure~3 and animations online). The first CME quickly
becomes flattened and distorted into a concave-outward shape,
presumably owing to the interaction with ambient structures and/or
solar wind speed gradient \citep{liu06a, liu08a, liu09b, savani10}.
A wave-like structure is observed ahead of the second CME, best seen
in HI1 of STEREO A, which may be ambient structures deflected by the
CME-driven shock or the shock itself. In situ measurements at 1 AU
do show a shock preceding the ejecta (see Figures~5 and 6).
Animations made of composite images (with FOVs to scale) are
available online, which show the evolution of the CMEs (and the
shock) in virtually the entire Sun-Earth space.

Note that the HI images require a special processing procedure prior
to the running differencing. This is necessary due to the large FOVs
and increasing faintness of CME signals as they move further from
the Sun. A background, computed from several days worth of data
before the events, is first subtracted from each image to remove the
F corona. We then align adjacent images before making the
running-difference sequence in an effort to eliminate the stellar
background. Finally, a median filter is applied to the
running-difference images to reduce the residual stellar effects.
Artifacts are still visible in Figure~2 (bottom two rows), including
the diffuse Milky Way galaxy in STEREO A and vertical streaks in
STEREO B (resulting from saturation due to planets).

A radial slit with a width of 64 pixels around the ecliptic plane is
extracted from the difference images of COR2, HI1 and HI2. Resistant
means of the running difference intensities, taken over the 64
pixels, are then stacked as a function of time and elongation, which
results in the time-elongation maps shown in Figure~3. The
elongation angles are plotted in a logarithmic scale to expand COR2
data. Tracks associated with the three CMEs and the shock can be
identified from the maps: the elongation angles along the tracks are
marked in the difference images to seek the corresponding structures
(see Figure~2). The shock driven by the second CME gives rise to the
fourth track as indicated in the maps of STEREO A; the remaining
three tracks are produced by the edges (mostly leading edges) of the
CMEs in the ecliptic plane. A smooth transition is observed in the
tracks from COR2 to HI1 and then from HI1 to HI2, indicative of a
continuous tracking of the same event over large distances. Also
shown is the elongation angle of the Earth, which is about
$84^{\circ}$ for STEREO A and $72^{\circ}$ for STEREO B.

Elongation angles of the CMEs (and the shock as well) can then be
extracted from the tracks, usually along the trailing edge of the
tracks (the black/white boundary) where the contrast is the
sharpest. When geometric triangulation is feasible with data from
both spacecraft, interpolation is performed to get elongation angles
at the same time tags for STEREO A and B as required by the
triangulation analysis; the values of the elongation angles are then
input to an appropriate set of equations to calculate the
propagation direction and radial distance (see \S 2.3). When data
are available from only one spacecraft, we fit the elongation angles
using equation~(5) assuming a constant propagation direction and
speed. Only HI1 and HI2 data are used in the track fitting. Table~1
lists the propagation direction, speed and predicted arrival time at
1 AU obtained from the track fitting. The propagation direction
($\beta_A$ or $\beta_B$) is converted to an angle with respect to
the Sun-Earth line. If the angle is positive (negative), the CME
feature would be propagating west (east) of the Sun-Earth line in
the ecliptic plane. The fits are also plotted in Figure~3 over the
time-elongation maps; a good agreement with the tracks is achieved.

Figure~4 shows the CME kinematics resulting from the triangulation
analysis and the comparison with track fitting. An uncertainty of 10
pixels, which is roughly 0.04$^{\circ}$, 0.2$^{\circ}$ and
0.7$^{\circ}$ for COR2, HI1 and HI2 respectively, is applied to the
measurements of elongation angles. This uncertainty does not
necessarily reflect the errors due to the effects of projection,
Thomson scattering and CME geometry, but is mainly to show the
sensitivity of the technique to the errors in the elongation angle
measurements (see Appendix A for error analysis). Note that the
diagram shown in the right panel of Figure~1 is invoked for the
first and third CMEs while the one shown on the left is used for the
second event. For the first and third CMEs, triangulation is applied
only to COR2 data; neither of the two events is observed by the HIs
of STEREO A since their FOVs are off to the east side of the Sun.
The second CME is not visible in HI2 of STEREO A (see Figures 2 and
3), so triangulation can only be performed out to HI1. The first CME
has a propagation direction increasing from about 60$^{\circ}$ to
73$^{\circ}$ west of the Sun-Earth line, comparable to the estimate
from track fitting; the radial distances connect well with the fit
ones, except that the fit speed is larger than that estimated from
geometric triangulation. Agreement between triangulation and fitting
is also seen for the third CME in terms of distance and speed, but
the propagation angle from track fitting seems larger, 106$^{\circ}$
compared with 93$^{\circ}$. The largest difference between
triangulation and fitting is observed for the second CME, with
1$^{\circ}$ versus $17^{\circ}$ for the propagation direction and
500 km s$^{-1}$ versus 390 km s$^{-1}$ for the speed (see Table~1).

The third CME is about $20^{\circ}$ west of the first one as
determined from geometric triangulation, which is almost the same
angle the Sun has rotated to the west during the time between the
launch of the two events. Therefore, these two CMEs are likely to
originate from the same source region on the Sun. The same
conclusion has been reached by \citet{ht08} in their geometrical
analysis of the two limb events. Only the second CME can be tracked
out to HI1 by the triangulation technique. Its propagation direction
changes from eastward to westward rapidly and then stays roughly
constant at $1^{\circ}$ west of the Sun-Earth line. This transition
is likely true: at the early stage STEREO A observed an elongation
larger than STEREO B (see the second row of Figure~2), but later the
elongations seen by the two spacecraft are similar (see the fourth
row of Figure~2). The speed of the second CME first increases and
then decreases into the FOV of HI1, which indicates a strong
interaction of the CME with the background heliosphere when not far
from the Sun. Also note that all the three CMEs (and the 2008
December 12 CME as well) undergo a westward migration at the
beginning. We suspect that this is a universal feature for all CMEs:
the strong magnetic field within CMEs, which is still connected to
the Sun as CMEs move outward, would produce a tendency for the
ejecta to co-rotate with the Sun (see discussion in \S 4.1).

The association with in situ signatures can now be established.
Figure~5 shows an MC identified from near-Earth solar wind data
based on the depressed proton temperature, strong magnetic field and
smooth rotation of the field. The magnetic field is measured in RTN
coordinates (in which \textbf{R} points from the Sun to the
spacecraft, \textbf{T} is parallel to the solar equatorial plane and
points to the planet motion direction, and \textbf{N} completes the
right-handed triad). A similar plasma and magnetic field structure
is observed at ACE and WIND. A preceding shock, as can be seen from
simultaneous increases in the plasma density, bulk speed,
temperature and magnetic field strength, passed the spacecraft at
17:17 UT on November 19. The density within the MC is comparable to
that of the ambient solar wind (upstream of the shock) but much
smaller than in the sheath (a transition layer between the MC front
and the shock) and trailing region. Also plotted are the predicted
arrival times of the four features estimated from track fitting (see
Table~1). The first CME is far earlier than the actual arrival while
the third one is much later; their propagation directions also seem
too westward to reach the Earth. Only the second CME shows the right
propagation direction and arrival time at 1 AU. Therefore, it must
be the second CME that is responsible for the MC. The predicted
shock arrival time is about 8 hr earlier than observed at the Earth.

The conclusion about the association is further supported by in situ
data from STEREO A and B as shown in Figure~6. Only a density spike
is observed at STEREO A, which may be produced by the disturbance
associated with the MC. STEREO B observed a shock at 13:48 UT on
November 19 and a subsequent MC on November 20. Presumably this is
the same event as observed near the Earth. The MC interval is mainly
determined from the strong magnetic field and rotation of the field.
Irregularities in the magnetic field are seen within the MC,
indicative of distortion by ambient structures. Enhanced
suprathermal electrons are observed during the time period but
mainly in a direction either parallel or anti-parallel to the
magnetic field (i.e., no bi-directional streaming). Again, only the
second CME shows the correct arrival time. The shock arrives at
STEREO B about 4 hr later than predicted by track fitting. Note that
the shock arrives at STEREO B before it arrives at the Earth
although STEREO B is slightly further from the Sun than the Earth.
This may indicate either a complex structure of the shock or a
change in the propagation direction owing to interactions with the
ambient medium. It is worth mentioning that the shock has a
substantially large standoff distance from the MC (compared with the
radial width of the MC which is about 0.08 AU); this seems
consistent with imaging observations from HI1 of STEREO A (see
Figure~2 and online animations).

Now that the correspondence has been established, we can compare the
structure determined from coronagraph image modeling with in situ
reconstruction. Figure~7 shows the observed and modeled images of
the second CME from three viewpoints. The fit is obtained by
adjusting the model parameters to match STEREO A and B images; LASCO
images are used to verify the fit. There are multiple structures in
LASCO with the most prominent one moving toward the northwest; a
very faint front on the left side, which is halo like and barely
seen in the still image, fits the model well. We actually tried to
fit the front on the northwest at first, but then a good match with
STEREO A and B images cannot be obtained simultaneously. The fit,
which is considered to match the three views, gives a propagation
direction about $2^{\circ}$ east of the Sun-Earth line and
$\pm1^{\circ}$ relative to the ecliptic plane as well as a rope tilt
angle about $-36^{\circ}$ (clockwise from the ecliptic; see
Figure~7). Whether the propagation direction is above or below the
ecliptic plane cannot be determined accurately from image modeling.
The image modeling suggests that the CME is headed almost right
toward the Earth, consistent with the geometric triangulation
analysis. Other CMEs are also simulated by the forward modeling
technique. Table~2 shows the resulting propagation direction and
rope orientation. The propagation angle of the 2007 November 14 CME
relative to the Sun-Earth line determined from image modeling is
similar to the estimate from geometric triangulation; for the 2007
November 16 CME, image modeling gives a larger angle but the
difference is not significant (123$^{\circ}$ versus 93$^{\circ}$).

Figure~8 displays the cross sections of the MC reconstructed from
the in situ data at ACE and STEREO B, respectively. The contours
represent nested helical magnetic field lines projected onto the
cross section in a flux-rope frame (with $x$ almost along the
spacecraft trajectory and $z$ in the direction of the axial field).
Table~3 gives the times, estimated axis orientations and magnetic
field chiralities for the MCs of interest. The field configuration
is left-handed at both ACE and STEREO B, as can be seen from the
transverse fields along the spacecraft trajectory. The in situ
reconstruction gives an axis elevation angle of about $-1.4^{\circ}$
in RTN coordinates at ACE while $-33.8^{\circ}$ at STEREO B. The
flux-rope tilt angle at STEREO B is comparable to the estimate from
image modeling ($-36^{\circ}$), but near the Earth it becomes
significantly smaller. The axis azimuthal angle is $106.7^{\circ}$
at ACE and $91.8^{\circ}$ at STEREO B (RTN), so the flux-rope axis
is nearly perpendicular to the radial direction (\textbf{R}), which
seems consistent with the scenario shown in Figure~7 (see the
simulated image for LASCO). The RTN directions are projected onto
the cross section in order to compare reconstruction results with
observations. For example, as ACE moves along $x$ in the flux-rope
frame, it would see a $B_R$ component that is first negative and
then positive, a $B_T$ that is largely positive (since the flux-rope
axis is almost along \textbf{T}), and a $B_N$ that is first positive
and then negative (see the field orientation along the spacecraft
trajectory). A similar magnetic field structure is observed at
STEREO B, but note a larger axis elevation angle than at ACE. These
results are consistent with the in situ measurements (see Figures~5
and 6).

The reconstructed cross sections at both ACE and STEREO B show a
maximum axial field below the ecliptic plane (as shown by the
spacecraft trajectory). The overall propagation direction of the CME
at 1 AU is thus likely to be southward. This can be easily
understood from the reconstruction results at ACE: the flux rope
axis is nearly parallel to the ecliptic plane and perpendicular to
the radial direction, and ACE crosses the MC above the flux-rope
axis, so the flux rope should be largely below the ecliptic plane.
The interpretation about the propagation direction is consistent
with the reconstruction at STEREO B although the flux rope is more
tilted there. Coronagraph image modeling of the CME indicates a
propagation direction within $\pm1^{\circ}$ of the ecliptic plane.
The propagation direction as well as the flux-rope orientation may
change during the transit from the Sun to 1 AU, possibly owing to
interactions with the background heliosphere \citep[also
see][]{liu08b}.

The Earth-directed event shows a few interesting while puzzling
features that deserve a further study. The CME is clearly visible in
HI2 of STEREO B but hardly discernible in HI2 of STEREO A (see
Figures 2 and 3), which indicates a propagation direction closer to
STEREO A than B (i.e., west of the Sun-Earth line). \citet{ht09}
obtain a central longitude of about $-17^{\circ}$ (i.e., almost
directly toward STEREO B) by fitting the CME as a spherical shell.
Our track fitting, geometric triangulation and image forward
modeling give a propagation angle of $17^{\circ}$, $1^{\circ}$ and
$-2^{\circ}$ (note the signs), respectively. The estimate of the CME
propagation direction by \citet{ht09} seems to contradict with the
HI observations but appears more or less consistent with the in situ
measurements, i.e., the ICME impacts the Earth and STEREO B but only
grazes STEREO A (see Figures 5 and 6). Note that a more regular
flux-rope structure is observed at the Earth than at STEREO B. As
mentioned earlier, the shock arrives at STEREO B before it arrives
at the Earth although STEREO B is 0.04 AU further from the Sun. A
high-speed flow is observed following the ICME (see Figures 5 and
6), which may squeeze the ejecta from behind. All these features
indicate a complex structure (other than spherical) and propagation
direction of the event due to interactions with the ambient flows.

\subsection{2008 December 12-17 Event}

The kinematics of the 2008 December 12 CME have been studied in
paper 1 with the geometric triangulation method. Here we briefly
summarize some of the results relevant to the present work. During
the time of the CME, the longitudinal separation between STEREO A
and B is about $\gamma\simeq86.3^{\circ}$, and the distances of the
two spacecraft from the Sun are $d_A \simeq 0.97$ AU and $d_B \simeq
1.04$ AU. The CME is associated with a prominence eruption in the
northern hemisphere. It produces two tracks in the time-elongation
maps, one corresponding to the leading edge of the CME and the other
the trailing edge, which extend out to 50$^{\circ}$ elongation for
both STEREO A and B. The triangulation analysis assuming $d_A = d_B
=1$ AU in paper 1 gives propagation directions generally within
$10^{\circ}$ of the Sun-Earth line, radial distances out to 150
solar radii or 0.7 AU, and predicted arrival times and radial speeds
consistent with the near-Earth in situ measurements around an MC.
Note that the CME also shows a westward migration when it is within
about 20 solar radii from the Sun. Refer to paper 1 for details and
animations of the imaging observations.

We repeat the analysis of paper 1 using the exact values of the
spacecraft distances and plot the results in Figure~9. Differences
are observed in the trends of the propagation angle but only at
large distances. The CME leading edge shows a continuous westward
deflection in the FOV of HI2, rather than suddenly turns to the east
of the Sun-Earth line as indicated in paper 1. The CME trailing edge
has a generally constant propagation angle with respect to the
Sun-Earth line in both HI1 and HI2, i.e., no turn to the east of the
Sun-Earth line. Other results, such as the radial distance and
speed, are almost the same as in paper 1 and consistent with the in
situ measurements around the MC. With the association between solar
observations and in situ signatures established by the triangulation
analysis, we can now compare the structures and propagation
directions determined from coronagraph image modeling and in situ
reconstruction.

Figure~10 shows the comparison between observed and modeled
coronagraph images of the 2008 December 12 CME from three
viewpoints. All the three views are reproduced fairly well; the
spatial extent of the simulated CME also agrees with the
observations, as shown by the wireframe rendering superposed on the
observed image. The fit from the forward modeling gives a
propagation direction about 10$^{\circ}$ west of the Sun-Earth line
and 8$^{\circ}$ above the ecliptic plane, and a flux-rope tilt angle
about $-53^{\circ}$ clockwise from the ecliptic plane (see Table~2).
The propagation angle relative to the Sun-Earth line determined from
image modeling is consistent with the estimate from the geometric
triangulation analysis. The basic structure of the CME is not
significantly distorted out to the FOV of HI1 (close to the Sunward
edge), as can seen in Figure~11; it is remarkable that at large
distances the CME can still be simulated by such a rope-like model.

Figure~12 shows the corresponding MC near the Earth identified from
the strong magnetic field and smooth rotation of the field. Again,
ACE and WIND observed a similar velocity and magnetic field
structure, but ACE does not have valid measurements of the proton
density and temperature due to the low solar wind speed. The
predicted arrival times of the CME leading and trailing edges are
good to within a few hours. Note that what is being tracked is
enhanced density regions outside the flux rope; the CME front has
swept up and merged with the ambient solar wind during its
propagation in the heliosphere. The density within the flux rope is
lower than that of the ambient solar wind owing to expansion, so the
flux rope probably cannot be imaged in white light at large
distances (especially in the FOV of HI2). Two small density spikes
are observed within the MC, reminiscent of the prominence material.
No ICME signatures are observed at STEREO A. STEREO B observed a
depressed proton temperature between 14:00 UT on December 16 and
02:00 UT on December 17, but at the same time the magnetic field is
not enhanced and does not show a coherent rotation. It is likely
that the ICME missed the two spacecraft given such a small event and
large spacecraft separation in longitude (or the ICME may have been
so well assimilated into the ambient structures that it is no longer
recognizable).

The MC cross section reconstructed from WIND data is displayed in
Figure~13. The transverse magnetic fields along the spacecraft
trajectory indicate a left-handed flux-rope configuration. The in
situ reconstruction gives a flux-rope tilt angle about
$-6.4^{\circ}$ and azimuthal angle about $94.9^{\circ}$ in RTN
coordinates (see Table~3). The flux-rope tilt angle is much smaller
than the estimate from image modeling ($-53^{\circ}$). The
reconstructed cross section shows a maximum axial field above the
ecliptic plane (as shown by the trajectory of WIND), so the overall
propagation direction of the CME at 1 AU is likely northward. This
is consistent with the results from image forward modeling. Again,
the RTN directions are projected onto the cross section in order to
compare reconstruction results with in situ measurements. As shown
by the field orientation along the spacecraft trajectory, the
magnetic field would have a $B_R$ component that is first positive
and then slightly negative, a $B_T$ that is largely positive (since
the flux-rope axis is almost along \textbf{T}), and a $B_N$ that is
first positive and then negative. This is exactly observed at WIND
and ACE (see Figure~12).

\citet{davis09} perform the same track fitting approach on the HI
observations of the event as in \S 3.1 and obtain similar results
(including propagation direction, speed and arrival time at 1 AU).
Refer to paper 1 for a comparison between the geometric
triangulation and track fitting for this event. \citet{lugaz10}
study the same event using only HI data with a similar triangulation
technique but assume the CME as a spherical front attached to the
Sun (see Appendix B). Their analysis yields a similar but somewhat
larger propagation angle. It should be noted that, although
\citet{lugaz10} argue their approach may be more appropriate than
our triangulation technique, they did not provide any comparison
with in situ measurements, which is the best means to test the
results (e.g., predicted arrival time and speed at 1 AU). Also see
Appendix B for a discussion of their method.

\section{Conclusions and Discussion}

We constrain the global structure and kinematics of CMEs by
combining image observations with in situ measurements. The global
structure of CMEs is reproduced from coronagraph observations by a
forward modeling technique, while the in situ counterpart at 1 AU is
reconstructed with the GS method. Propagation of CMEs between the
Sun and 1 AU is studied with a geometric triangulation technique,
which enables a proper association between imaging and in situ
observations.

In addition to the case studies described in \S 3, we are also
performing a statistical analysis of Earth-directed events with
joint imaging and in situ observations. All the results, including
movies made of composite images, time-elongation maps, CME
kinematics derived from triangulation analysis, plots showing in
situ signatures and comparison with triangulation analysis, and in
situ reconstruction if possible, are compiled into our catalog of
STEREO Earth-directed CMEs at
\url{http://sprg.ssl.berkeley.edu/~liuxying/CME_catalog.htm}. One
focus of the statistical study is on the evaluation of the geometric
triangulation technique. Our preliminary results from the
statistical study show that the CME arrival time and speed at the
Earth are generally well predicted by the triangulation method. All
the CMEs studied so far exhibit more or less a westward motion at
their acceleration phase, similar to the 2007 November and 2008
December events. Below we will summarize and discuss the results
mainly based on the case studies described in \S 3.

\subsection{Consistency and Caveats}

Imaging observations and in situ measurements are two basic means in
probing CME properties. They are essentially distinct in terms of
the physics relied on to do the measurements. The following findings
from this work may help pave the way to link imaging and in situ
observations, as required by the proper treatment of the CME problem
and practical space weather forecasting.

First, CME propagation directions derived from different methods are
generally consistent with each other. For the 2007 November 15 and
2008 December 12 CMEs, both the geometric triangulation analysis and
coronagraph image modeling result in propagation directions almost
right toward the Earth; in situ measurements near the Earth show
corresponding signatures with the correct timing. A rough agreement
between triangulation analysis and image modeling is also obtained
for the 2007 November 14 and 16 events which propagate west of
STEREO A and B; the possibility for them to reach the Earth is
excluded based on the timing and their propagation directions. In
situ reconstruction of the 2008 December 12 event at 1 AU shows a
maximum axial magnetic field above the ecliptic plane, which agrees
with the overall northward propagation obtained from coronagraph
image modeling. For the 2007 November 15 CME, in situ reconstruction
at 1 AU indicates a southward propagation, but a rigorous comparison
with coronagraph image modeling is not possible.

Second, the geometric triangulation technique shows a promising
capability to link solar observations with corresponding in situ
signatures at 1 AU and to predict CME arrival at the Earth.
Association between solar observations and in situ signatures at 1
AU is often ambiguous due to the large distance gap; the situation
becomes worse for consecutive CMEs like the 2007 November events.
This may be the main reason that most CME studies are focused only
on the Sun or ICME signatures near the Earth. The geometric
triangulation technique, which can determine both propagation
direction and radial distance continuously over an extensive region
of the heliosphere and thus arrival time at 1 AU, provides a means
to identify the unique association between solar observations and in
situ signatures. It should be stressed that, even though the effect
of CME geometry is not taken into account, the technique still
presents a reasonable accuracy in determining CME kinematics and
arrival time as shown by the comparison with in situ measurements
and coronagraph image modeling.

Third, the flux rope within CMEs probably cannot be imaged at large
distances by the HIs due to expansion. The tracks that extend to
large elongations in time-elongation maps are usually CME edges, so
what is being tracked is enhanced density regions outside the flux
rope. This finding is confirmed by in situ measurements. On average,
the plasma density within the flux rope at 1 AU is comparable to the
ambient density, i.e., about 5 cm$^{-3}$ \citep[see][and Figures 5,
6 and 12]{liu06b, liu06c}. A correlation study of solar wind stream
interaction regions between HI2 images and in situ measurements at 1
AU seems to suggest a density threshold of about 15 cm$^{-3}$ at 1
AU for HI2 observations \citep{sheeley08}. Therefore, the density
within the flux rope is usually far below the HI2 sensitivity. This
raises a serious problem for space weather forecasting: the arrival
time of the flux rope, which has the most hazardous southward
magnetic field, cannot be predicted precisely from imaging
observations. The uncertainty could be as large as 10 hr depending
on the size of the sheath region; the average duration of the sheath
at 1 AU is about 14 hr when the ejecta is preceded by a shock
\citep{liu06c}. However, it is usually the shock that brings the
sudden commencement of geomagnetic storms; the sheath region is also
geoeffective although the magnetic field is turbulent there
\citep{liu08a}.

Fourth, the flux-rope orientation derived from in situ
reconstruction at 1 AU may have a large deviation from that
determined by coronagraph image modeling. The image modeling of the
2007 November 15 CME gives a tilt angle of about $-36^{\circ}$, much
larger than the estimate from in situ reconstruction near the Earth
($-1.4^{\circ}$), although a similar result is obtain from the in
situ data at STEREO B. For the 2008 December 12 CME, image modeling
and in situ reconstruction yield a flux-rope tilt angle about
$-53^{\circ}$ and $-6.4^{\circ}$ respectively, an even larger
discrepancy. These results seem contrary to the suggestion of
\citet{rouillard09}, which is based on a single case study, that the
flux-rope orientation can be predicted from white-light images. The
lack of consistency with in situ measurements is likely accounted
for by the evolution of the flux rope in interplanetary space, e.g.,
distortion/deflection by the ambient structures and/or flux rope
rotation between the Sun and 1 AU. As indicated by in situ
measurements, the density structures that appear to be part of a CME
in imaging observations are actually different regions from the flux
rope. The envelope fit by the image forward modeling might not
reflect precisely the exact size and location of the flux rope in
coronagraph observations. Faraday rotation measurements of the CME
magnetic field using polarized radio signals \citep{liu07} are thus
extremely valuable given the limited capability of imaging
observations in determining the magnetic field orientation.

Fifth, the triangulation analysis indicates that all the CMEs
studied here undergo a westward migration with respect to the
Sun-Earth line at their early stage. The 2007 November 14, 2007
November 15 and 2008 December 12 CMEs move westward by about
$13^{\circ}$, $4^{\circ}$ and $10^{\circ}$, respectively; the 2007
November 16 event does not achieve a radial motion in the FOV of
COR2 even after it has moved to the west by $8^{\circ}$. These
rotation angles should be considered as the lower limits since COR1
observations are not taken into account. The westward motion occurs
mainly when CMEs are accelerated (see Figures~4 and 9). Other events
in our CME catalog also show more or less a westward motion at their
acceleration phase. If this were an effect of CME geometry, then
there would be no systematic westward motion, i.e., the change in
the propagation angle would be randomly distributed between eastward
and westward, which is apparently not the case. Also note that, with
the two views, the propagation angle and distance are linearly
independent (see equation~4). To the best of our knowledge, this is
the first time that a systematic westward migration of CMEs relative
to the Sun-Earth line is discovered at the acceleration phase. A
direct consequence of the westward transition is that all techniques
for tracking CMEs, which assume a radial propagation from the solar
source region, may lead to a considerable error in the propagation
direction. Due to this westward transition, CMEs that occur at the
eastern hemisphere of the Sun would have a greater potential to
reach the Earth than events from the western hemisphere.

The westward motion, which is expected to be a universal feature for
CMEs at the early stage, can be explained by the magnetic field
connecting the Sun and CMEs. As the magnetic field is frozen in the
CME plasma, the Sun and a CME would be coupled together by the
magnetic field out to a distance, the so-called Alfv\'{e}n radius
$r_A$, within which the flow energy is dominated by the magnetic
field energy, i.e., $\rho v^2/2 \leq B^2/2\mu_0$. Here $\rho$, $v$
and $B$ are the mass density, speed and magnetic field strength of
the CME. Therefore, the magnetic field inside CMEs produces a
tendency toward co-rotation with the Sun. More specifically, the
westward migration of CMEs with respect to the Sun-Earth line is
caused by the rotation of the Sun when the motion of CMEs is
controlled by the magnetic field. The value of $r_A$ or co-rotation
angle is determined by the interplay between the density, magnetic
field and acceleration of CMEs. The CMEs examined here show an
Alfv\'{e}n radius of 10-20 solar radii (the distance at which CMEs
stop moving westward). This value seems larger than the counterpart
of the solar wind, which is usually below 10 solar radii but depends
on the solar wind source location, e.g., streamer belt, coronal
holes and active regions. The westward motion of CMEs at the early
stage will be further investigated in a separate paper.

\subsection{Concept for Future Missions}

This merged imaging and in situ study demonstrates the exciting
possibility to predict CME interplanetary properties with
observations from multiple vantage points. In particular, the
geometric triangulation concept based on wide-angle imaging
observations from STEREO can track CMEs (both propagation direction
and radial distance) continuously from the Sun all the way out to
the Earth. The same concept can be applied to future missions at the
fourth and fifth Lagrangian points (L4 and L5), which are well
situated for this purpose.

Figure~14 shows the five Lagrangian points in the Sun-Earth system.
These are fixed positions in an orbital configuration where a small
object (such as a spacecraft) can be theoretically stationary. L4
and L5 have the same orbit as the Earth but lie at $60^{\circ}$
ahead of and behind the Earth, respectively. Unlike STEREO A and B,
they are fixed in space with respect to the Sun and Earth rather
than drift away from each other. The longitudinal separation
($120^{\circ}$) between these two points is appropriate for the
observation of CMEs, even wide ones. Another advantage of L4 and L5
is that they are resistant to gravitational perturbations, so they
are truly stable. (L1, L2 and L3 are only meta stable; a spacecraft
at these points has to use frequent propulsions to remain on the
same orbit.) It would be of great merit to have dedicated spacecraft
making routine observations at both L4 and L5. They are located well
away from the Sun-Earth line, thus advantageous for observing
Earth-directed CMEs; triangulation with two such spacecraft makes it
possible to unambiguously derive the true path and velocity of CMEs;
we would also be able to determine the global structure and how the
Earth cuts through the structure. A hypothesized Earth-directed CME
is shown in Figure~14, illustrating the exciting possibility that
the interplanetary properties of the event can be accurately
determined well before it impacts the Earth. This observational
concept would be extremely important for CME research and represent
a major step toward practical space weather forecasting.

\acknowledgments The research was supported by the STEREO project
under grant NAS5-03131. Y. Liu thanks N. Lugaz of University of
Hawaii for helpful discussion. SECCHI was developed by a consortium
of NRL, LMSAL and GSFC (US), RAL and Univ. Birmingham (UK), MPI
(Germany), CSL (Belgium), and IOTA and IAS (France). We also
acknowledge the use of data from WIND, ACE and SOHO. R. Lin has been
supported in part by the WCU grant (No. R31-10016) funded by KMEST.

\appendix

\section{Error Analysis}

The triangulation technique presented in \S 2.3 allows an easy
evaluation of errors in the propagation direction and radial
distance due to uncertainties in elongation measurements. The two
spacecraft make independent measurements of elongation angles, so
the errors can be expressed as
\begin{equation}
\sigma_{\beta} = \sqrt{\left(\frac{\partial \beta_A}{\partial
\alpha_A}\right)^2\sigma^2_{\alpha A} + \left(\frac{\partial
\beta_A}{\partial \alpha_B}\right)^2\sigma^2_{\alpha B}},
\end{equation}
\begin{equation}
\sigma_{r} = \sqrt{\left(\frac{\partial r}{\partial
\alpha_A}\right)^2\sigma^2_{\alpha A} + \left(\frac{\partial
r}{\partial \alpha_B}\right)^2\sigma^2_{\alpha B}},
\end{equation}
for the propagation direction and radial distance, respectively,
where $\sigma_{\alpha}$ represents uncertainties in the elongation
measurement. For a CME propagating between the two spacecraft,
equations (1) and (4a) result in
\begin{equation}
\frac{\partial \beta_A}{\partial \alpha_A} =
2f\sin\alpha_B\left[\sin(\alpha_B + \gamma) -
f\sin\alpha_B\right]/X,
\end{equation}
\begin{equation}
\frac{\partial \beta_A}{\partial \alpha_B} =
2\sin\alpha_A\left[\sin\alpha_A - f\sin(\alpha_A + \gamma)\right]/X,
\end{equation}
\begin{equation}
\frac{\partial r}{\partial \alpha_A} = \frac{d_A}{\sin(\alpha_A +
\beta_A)}\left[\cos\alpha_A - \frac{Y}{X}\sin\alpha_A\cot(\alpha_A +
\beta_A)\right],
\end{equation}
\begin{equation}
\frac{\partial r}{\partial \alpha_B} = \frac{-2d_A\cos(\alpha_A +
\beta_A)\sin^2\alpha_A[\sin\alpha_A -f\sin(\alpha_A +
\gamma)]}{X\sin^2(\alpha_A + \beta_A)},
\end{equation}
where $$Y = 1 - \cos(2\alpha_A) + 2f\sin\alpha_B\sin(2\alpha_A +
\alpha_B + \gamma),$$ $$X = 1 + f^2 - \cos(2\alpha_A) -
f^2\cos(2\alpha_B) + 4f\sin\alpha_A\sin\alpha_B\cos(\alpha_A +
\alpha_B + \gamma).$$ Similarly, expressions can be obtained for
CMEs propagating outside the space between the two spacecraft.
Consider an uncertainty of 10 pixels in the elongation angles
(corresponding to 0.04$^{\circ}$, 0.2$^{\circ}$ and 0.7$^{\circ}$
for COR2, HI1 and HI2 respectively) for the 2008 December 12 CME
which is propagating almost along the Sun-Earth line. The above
equations give typical errors of $0.5^{\circ}$, $1^{\circ}$ and
$3^{\circ}$ in the propagation direction and 0.07, 0.6 and 1.4 solar
radii in the radial distance for COR2, HI1 and HI2 respectively (see
Figure~9). Also see Figure~4 for CMEs west of the two spacecraft.

\section{Converting Elongation to Distance}

Coronagraphs and heliospheric imagers measure the elongation angles
of CMEs, not radial distances. The combined effects due to
projection, Thomson scattering and CME geometry form a major
challenge in determining CME kinematics. Here we summarize various
approximations that have been made to convert elongation to distance
and discuss their advantages and restrictions.

\subsection{Point P}

The point P (PP) approximation assumes a CME as a spherical front
centered at the Sun \citep[e.g.,][]{houminer72}. The brightest part
of the CME to a spacecraft is the region (point P) where the CME
intersects with the ``Thomson surface", a spherical front with the
Sun-spacecraft line as the diameter which has the maximum Thomson
scattering strength \citep{vourlidas06}, as illustrated in Figure~15
(left). The radial distance can be obtained from
\begin{equation}
r_{PP} = d\sin\alpha.
\end{equation}
This is the simplest way to convert elongation measurements to
radial distances while taking into account Thomson scattering
effects. Apparently, the CME geometry is oversimplified, and as a
result the technique is only applicable to extremely wide events.
Information about the propagation direction cannot be obtained. The
method provides a lower limit of the distance (see below).

\subsection{Fixed $\beta$}

The fixed $\beta$ \citep[F$\beta$; or fixed $\phi$ using the
terminology of][]{kahler07} approximation takes the opposite
philosophy and assumes a relatively compact structure moving along a
fixed radial direction, as shown in Figure~15 (middle). It is first
proposed by \citet{sheeley99} and has been extensively used in track
fitting (see \S 2.3). The F$\beta$ method results in
\begin{equation}
r_{F\beta} = \frac{d\sin\alpha}{\sin(\alpha+\beta)},
\end{equation}
which has been adopted for geometric triangulation in \S 2.3. The
advantage of this technique is that it can provide the propagation
direction (say, through track fitting). It does not take into
account the effects of CME geometry and Thomson scattering, but
surprisingly it achieves a reasonable accuracy in both track fitting
and geometric triangulation (see \S 3 and paper 1). The assumption
of constant propagation direction is usually not true when CMEs are
close to the Sun (see \S 3); even far away from the Sun, they can be
deflected by ambient solar wind structures. Note that, when the
formula is used for geometric triangulation, the restriction to only
narrow CMEs can be relaxed; the assumption of fixed radial
trajectory is certainly abolished too.

\subsection{Harmonic Mean}

Given the restrictions of the PP and F$\beta$ approximations, a
compromise would be to take the harmonic mean (HM) of these two
formulas \citep{lugaz09}, namely, $$\frac{1}{r_{HM}} =
\frac{1}{2}\left(\frac{1}{r_{PP}} + \frac{1}{r_{F\phi}}\right),$$
which gives
\begin{equation}
r_{HM} = \frac{2d\sin\alpha}{1+\sin(\alpha+\beta)}.
\end{equation}
The physics behind this approximation is that CMEs are assumed as a
spherical front attached to the Sun moving along a fixed radial
direction; what is seen by the spacecraft is the segment tangent to
the line of sight, as shown by Figure~15 (right). It can be easily
proved that this CME geometry yields exactly the HM formula. Note
that the angles marked as $\epsilon$ are the same. The distances of
$r_{PP}$ and $r_{F\beta}$ are also indicated in the right panel of
Figure~15. Apparently, the PP approximation gives the shortest
distance for a given elongation angle while $r_{HM}$ (the diameter
of the sphere) lies in between. Inverting equation~(B3) for $\alpha$
yields
\begin{equation}
\alpha =
\arcsin\left(\frac{r_{HM}}{\sqrt{4d^2-4dr_{HM}\cos\beta+r^2_{HM}}}\right)
+ \arctan\left(\frac{r_{HM}\sin\beta}{2d-r_{HM}\cos\beta}\right).
\end{equation}
A similar track fitting procedure can be applied using this new
equation, but it is more complicated than equation~5, which may
diminish its efficiency. The HM approximation is intended for wide
CMEs only. It is argued that this technique can give better results
than the PP and F$\beta$ methods \citep{lugaz09}. To be fair,
restrictions also exist. CMEs can be significantly distorted by
ambient coronal and solar wind structures \citep{liu06a, liu08a,
liu09b}; even a uniform solar wind can easily flatten the cross
section with its radial flow \citep{riley04, liu06a}. Therefore, it
is very difficult for CMEs to maintain a spherical shape. The
assumption of a spherical front may be appropriate if the spacecraft
lies in the plane of the CME loop, but in principle the flux rope
can have any orientation.

\subsection{Triangulation with Harmonic Mean}

Motivated by the triangulation concept of \citet{liu10},
\citet[][and the present author himself]{lugaz10} realize that the
same idea can be applied with equation~(B3). Figure~16 shows the
diagram, which is similar to that in Figure~1 but assumes a
spherical front attached to the Sun. Each spacecraft observes
different plasma parcels which are assumed to be tangent to the line
of sight. Equations~(1) and (2) now become
\begin{equation}
\frac{r[1+\sin(\alpha_A+\beta_A)]}{2\sin\alpha_A} = d_A,
\end{equation}
\begin{equation}
\frac{r[1+\sin(\alpha_B+\beta_B)]}{2\sin\alpha_B} = d_B.
\end{equation}
Equation~(3) remains the same. This new triangulation concept might
be more appropriate than the original one, if the whole flux rope
lies in the ecliptic plane which is, however, not necessarily the
case. As discussed above, CMEs can hardly maintain a spherical
shape. Another advantage of the new method is that a CME can be seen
by the HIs on both spacecraft even if its nose is outside the two
spacecraft. This strength is accompanied by a dark side too. A rough
knowledge of where the CME nose is with respect to the two
spacecraft is needed apriori to determine which equation should be
used (3a, 3b, or 3c). This is difficult due to projection effects,
since the method assumes that what is observed is the region tangent
to the line of sight, not necessarily the nose. Another complication
is that the above equations have multiple solutions. In practice,
one needs to determine which solution should be adopted. A concern
is that, in some cases, the sines and cosines of the elongation and
spacecraft separation angles and their combinations are such that we
are not able to pick the right one.

It is immediately evident from Figure~16 why the original
triangulation concept gives a reasonable accuracy even when the two
spacecraft do not observe the same part of a CME: as long as the
propagation direction is not very far away from the Sun-Earth line,
the regions seen by the two spacecraft are not much deviated from
the nose. In addition, many narrow CMEs exist, and for those events
the original triangulation technique is more suitable than the new
one; it works even if those CMEs are outside the space between the
two spacecraft (see \S 3). The new triangulation notion is probably
complementary to, rather than supercedes, the original concept. It
would be helpful to have a statistical study with joint imaging and
in situ data and comparison with realistic MHD simulations to assess
the efficiency, simplicity and ease of use of these two
triangulation techniques.

\clearpage

\begin{table}
\caption{Estimated parameters of the four features from track
fitting}
\begin{tabular}{ccccc}
\tableline\tableline Feature & Spacecraft\tablenotemark{a} &
Direction\tablenotemark{b} ($^{\circ}$) & Speed
(km s$^{-1}$) & Arrival time at 1 AU (UT)\\
\tableline
1 &STEREO B &74    &449 &Nov 18, 20:14 \\
2 &STEREO B &17    &388 &Nov 20, 07:50 \\
3 &STEREO B &106   &394 &Nov 20, 22:01 \\
4 &STEREO A &$-29$ &474 &Nov 19, 09:28 \\
\tableline
\end{tabular}
\tablenotetext{a}{Spacecraft used for the track fitting; only HI1
and HI2 data are adopted.} \tablenotetext{b}{Propagation direction
with respect to the Sun-Earth line; positive if west and negative if
east.}
\end{table}

\clearpage

\begin{table}
\caption{CME parameters estimated from image modeling}
\begin{tabular}{lcccc}
\tableline\tableline Event & Time\tablenotemark{a} (UT) &
$\theta$\tablenotemark{b} ($^{\circ}$) & $\phi$\tablenotemark{c}
($^{\circ}$) & $\Theta$\tablenotemark{d} ($^{\circ}$) \\
\tableline
2007 Nov 14 (feature 1) &Nov 15, 01:22 &$-19$  &61   &$-18$ \\
2007 Nov 15 (feature 2) &Nov 16, 01:22 &$\pm1$ &$-2$ &$-36$ \\
2007 Nov 16 (feature 3) &Nov 16, 16:22 &$-7$   &123  &1 \\
2008 Dec 12             &Dec 12, 12:52 &8      &10   &$-53$ \\
\tableline
\end{tabular}
\tablenotetext{a}{The time of the CME image being simulated.}
\tablenotetext{b}{Propagation direction with respect to the ecliptic
plane; positive if northward and negative if southward.}
\tablenotetext{c}{Propagation direction with respect to the
Sun-Earth line; positive if west and negative if east.}
\tablenotetext{d}{Tilt angle of the rope relative to the ecliptic
plane; positive if counterclockwise from the ecliptic and negative
if clockwise.}
\end{table}

\clearpage

\begin{table}
\caption{Estimated parameters of MCs at different spacecraft}
\begin{footnotesize}
\begin{tabular}{cccccccc}
\tableline\tableline Event & Spacecraft & Shock & Start & End &
$\Theta$\tablenotemark{a} & $\Phi$\tablenotemark{a} & Chirality \\
& & (UT) & (UT) & (UT) & ($^{\circ}$) & ($^{\circ}$) & \\
\tableline
2007 Nov &ACE/WIND &Nov 19, 17:17 &Nov 19, 23:17 &Nov 20, 11:17 &$-1.4$  &106.7 &L \\
         &STEREO B &Nov 19, 13:48 &Nov 19, 23:02 &Nov 20, 06:14 &$-33.8$ &91.8  &L \\
2008 Dec &ACE/WIND &-             &Dec 17, 03:36 &Dec 17, 14:38 &$-6.4$  &94.9  &L \\
\tableline
\end{tabular}
\end{footnotesize}
\tablenotetext{a}{Axis elevation and azimuthal angles in RTN
coordinates, respectively.}
\end{table}

\clearpage

\begin{figure}
\epsscale{1} \plotone{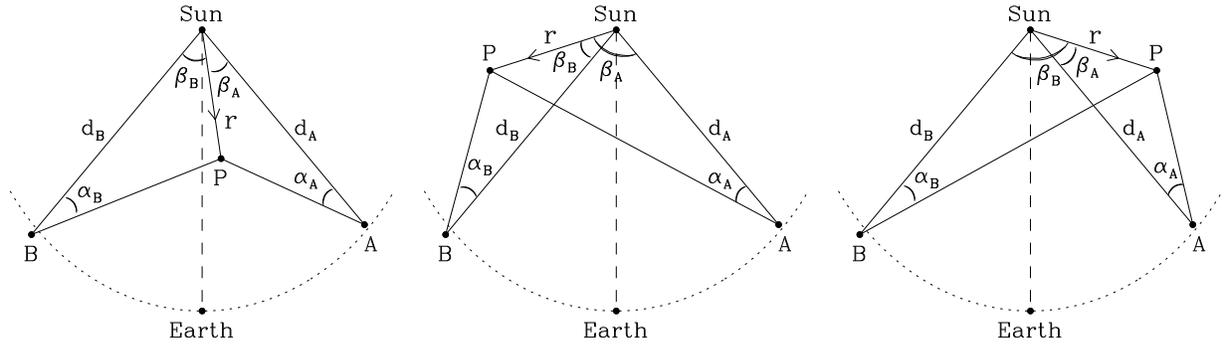} \caption{Diagrams of the geometric
triangulation in the ecliptic plane for a CME feature propagating
between (left), east of (middle), and west of (right) STEREO A and
B. The white-light feature is denoted by the point P with its
direction shown by the arrow. The dotted line indicates the orbit of
the Earth, and the dashed line represents the Sun-Earth line.}
\end{figure}

\clearpage

\begin{figure}
\epsscale{0.45} \plotone{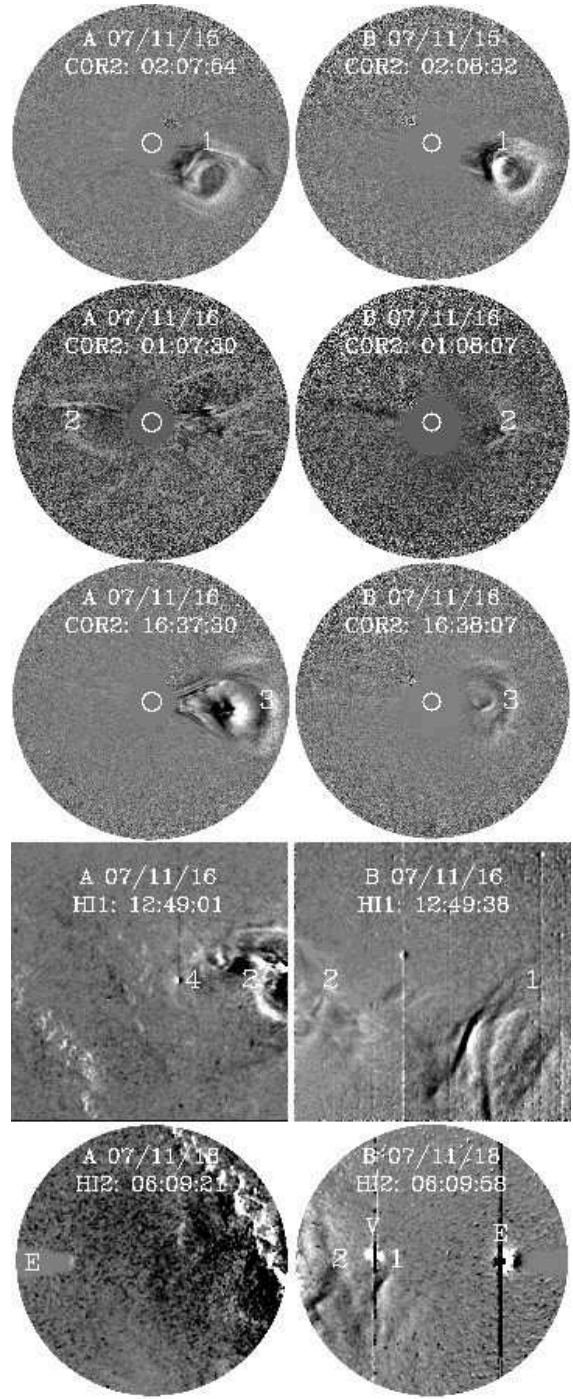} \caption{Running difference images
of the 2007 November CMEs observed by STEREO A (left) and B (right)
near simultaneously. The top three rows display COR2 images of the
three CMEs, respectively, and the bottom two rows show images from
HI1 and HI2. The numbers, corresponding to the tracks in Figure~3,
mark the locations (obtained from Figure~3) of the features. The
positions of the Earth and Venus are labeled as E and V. The Earth
is in the trapezoidal zone of STEREO A (the Earth occulter).
(Animations of this figure are available in the online journal.)}
\end{figure}

\clearpage

\begin{figure}
\epsscale{0.8} \plotone{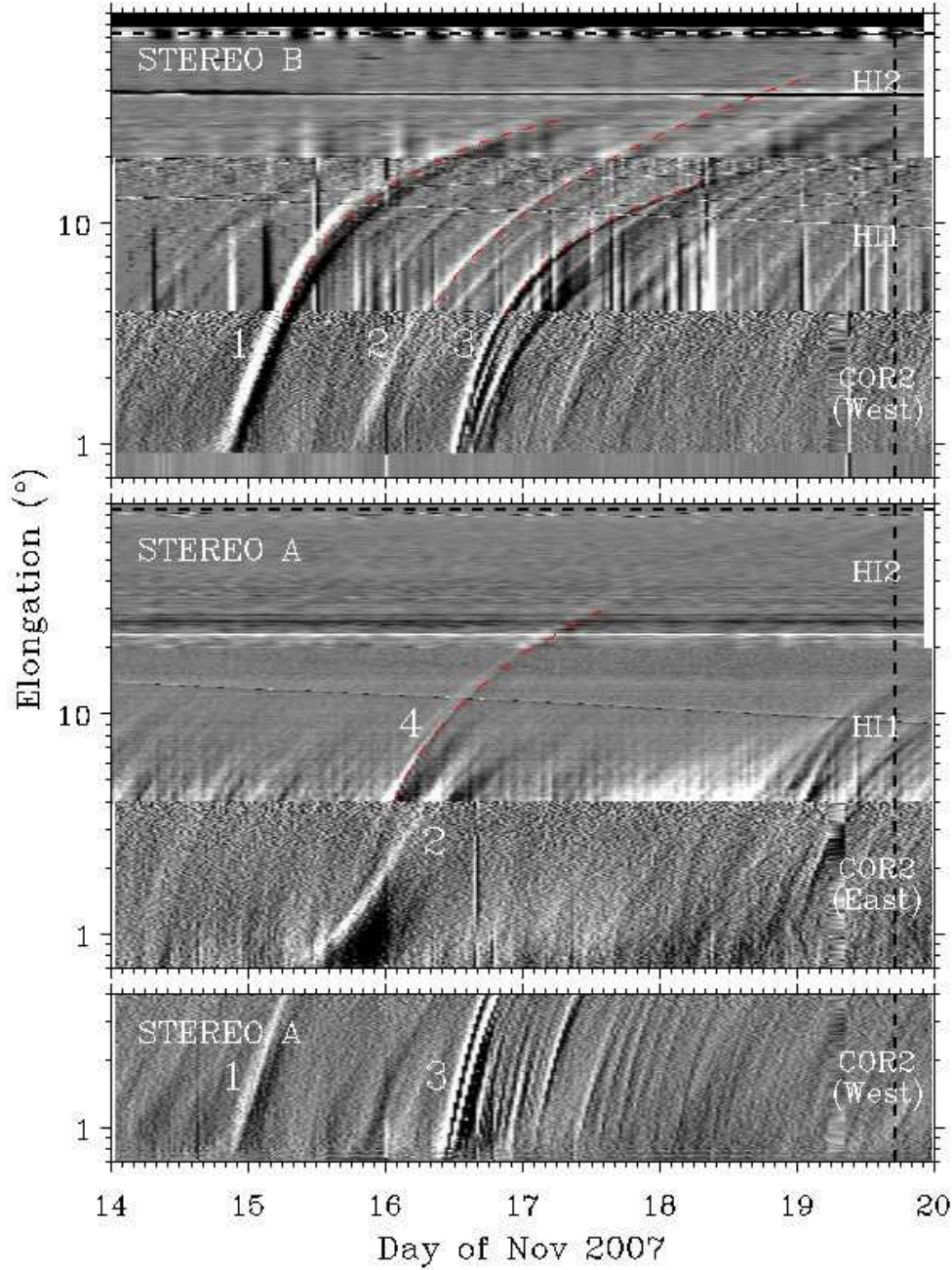} \caption{Time-elongation maps
constructed from running difference images of COR2, HI1 and HI2
along the ecliptic plane for STEREO A and B. Data from the western
part of COR2 (90$^{\circ}$ clockwise from the ecliptic north) are
also shown for STEREO A, as required by the triangulation analysis
for CMEs propagating west of the two spacecraft (see Figure~1). The
numbers indicate four tracks associated with the CMEs. The red
curves show the fits to the tracks in HI1 and HI2. The vertical
dashed line indicates the arrival time of a CME-driven shock at the
Earth, and the horizontal dashed line marks the elongation angle of
the Earth.}
\end{figure}

\clearpage

\begin{figure}
\epsscale{0.7} \plotone{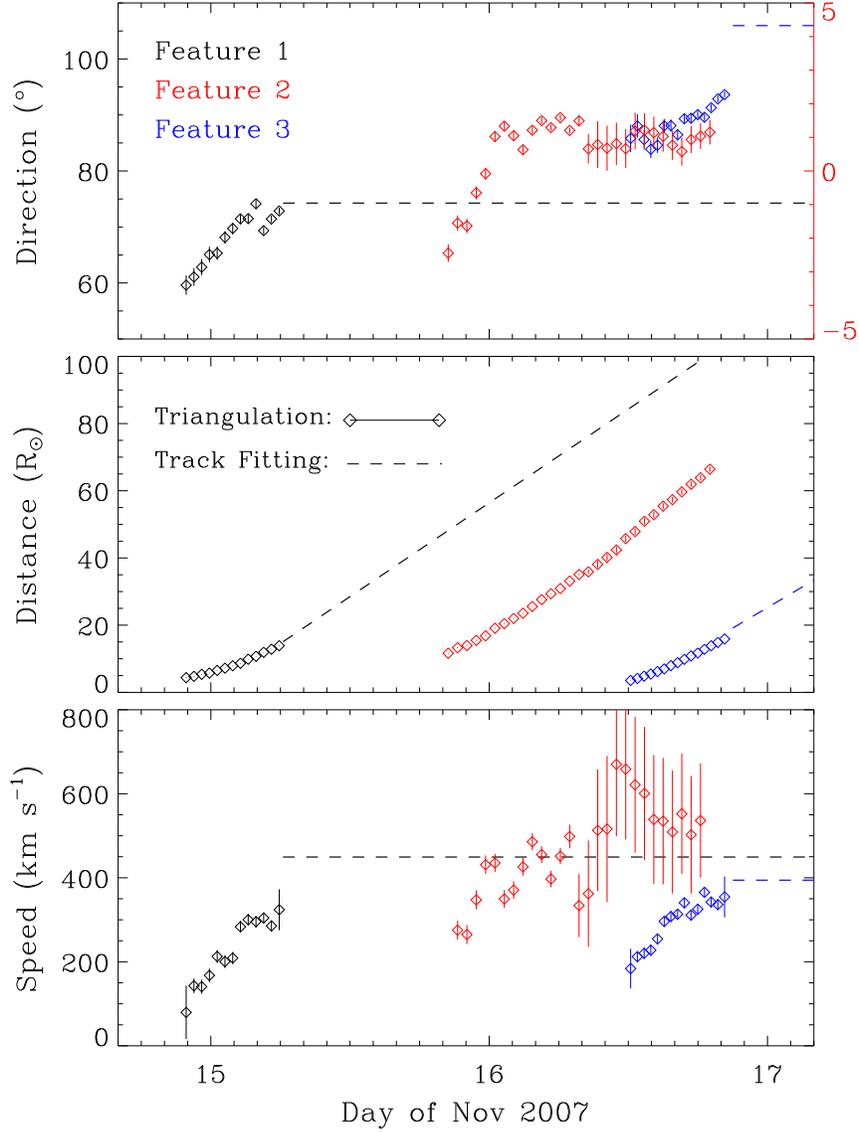} \caption{Propagation direction,
radial distance and speed of features 1 (black), 2 (red) and 3
(blue) derived from geometric triangulation (scatter) and track
fitting (dashed lines). For feature 2, the propagation direction is
scaled by the right axis (red). Track fitting results for feature 2
are not plotted here (but see Table~1). The speeds (scatter) are
calculated from adjacent distances using a numerical differentiation
with three-point Lagrangian interpolation. Error bars represent
uncertainties mathematically derived from the measurements of
elongation angles.}
\end{figure}

\clearpage

\begin{figure}
\epsscale{0.7} \plotone{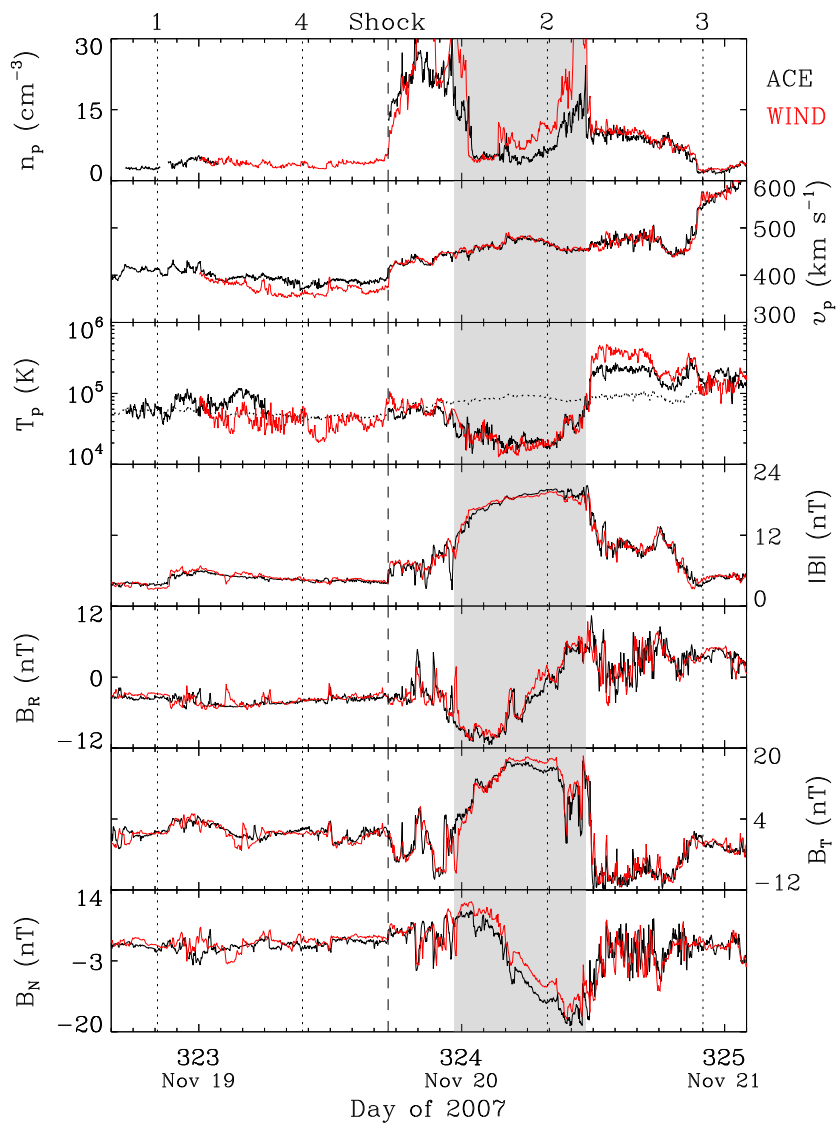} \caption{Solar wind plasma and
magnetic field parameters across the MC observed by ACE (black) and
WIND (red). From top to bottom, the panels show the proton density,
bulk speed, proton temperature, and magnetic field strength and
components. The shaded region indicates the MC interval. The
MC-driven shock and the predicted arrival times of the four features
at 1 AU are marked by the vertical dashed and dotted lines,
respectively. The dotted curve in the third panel denotes the
expected proton temperature from the observed speed.}
\end{figure}

\clearpage

\begin{figure}
\centerline{\includegraphics[width=20pc]{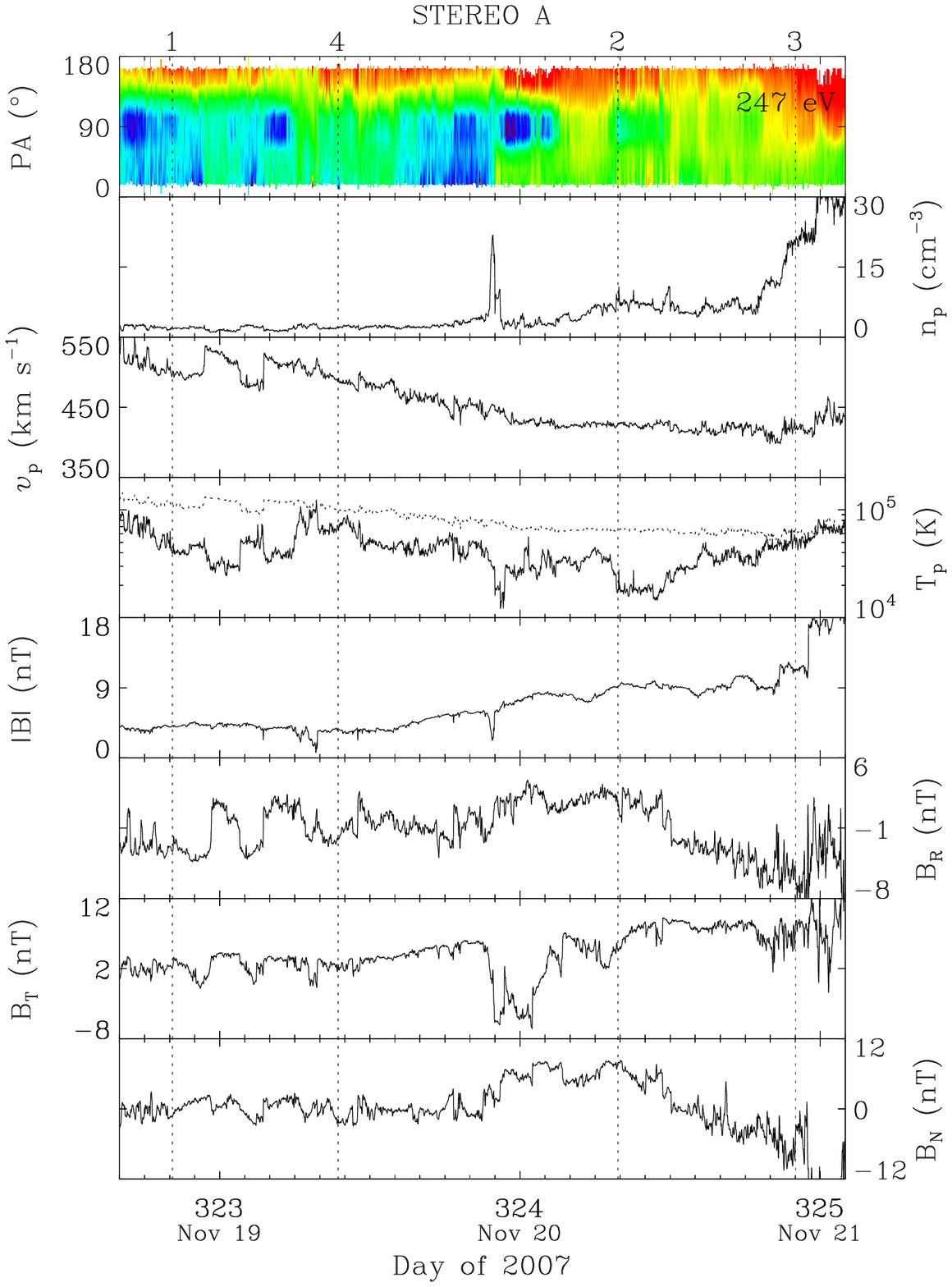}
\includegraphics[width=20pc]{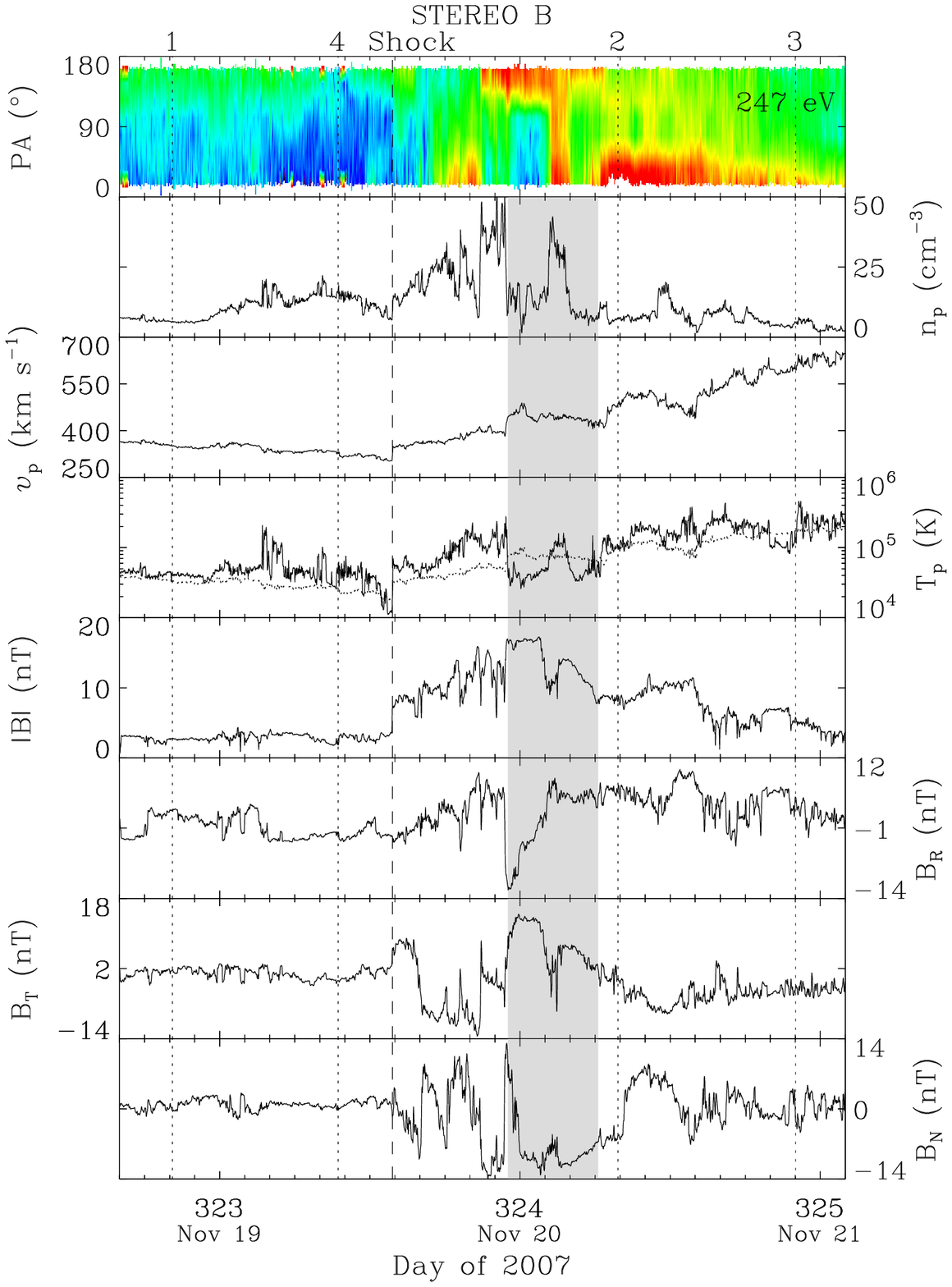}}
\caption{Similar format to Figure~5, but for the measurements at
STEREO A (left) and B (right). Also shown is the pitch-angle (PA)
distribution of 247 eV electrons (top). The color shading indicates
the values of the electron flux (descending from red to blue).}
\end{figure}

\clearpage

\begin{figure}
\epsscale{0.8} \plotone{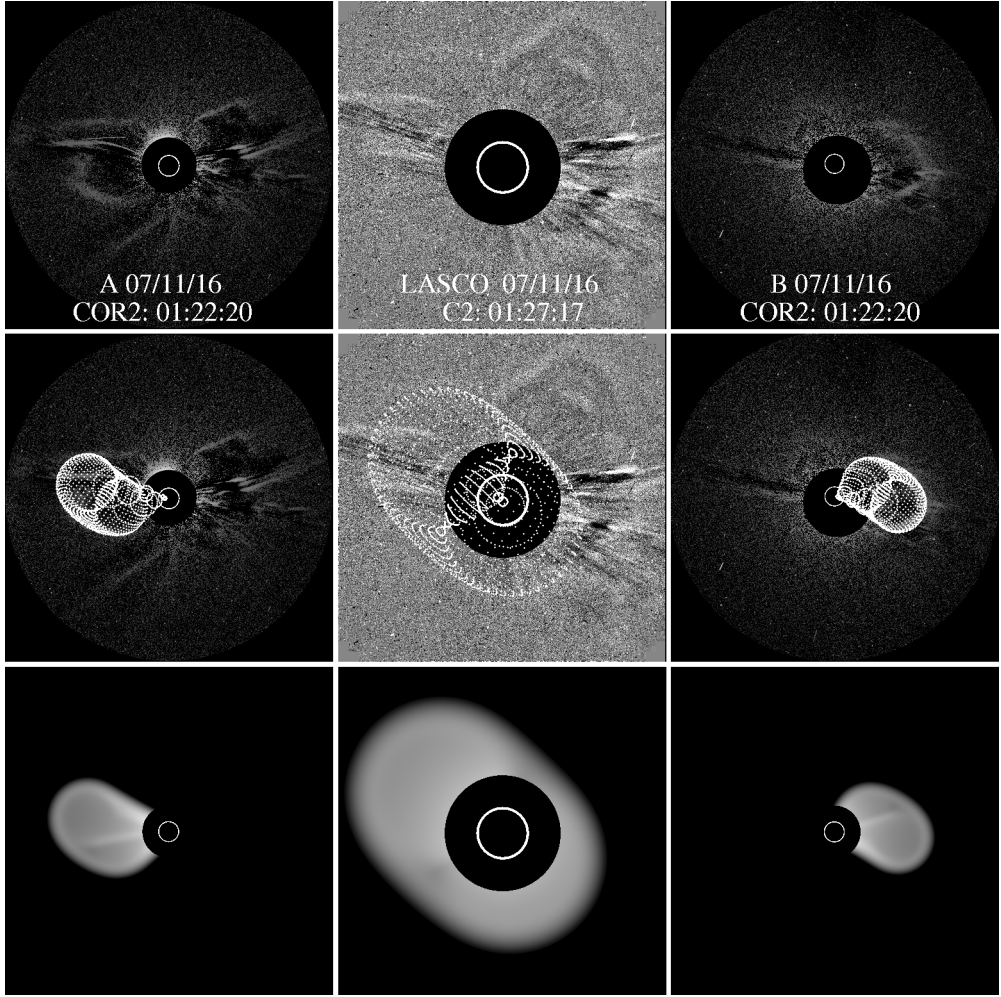} \caption{Observed (top) and modeled
(middle and bottom) images of the second CME as viewed from three
spacecraft. The middle panels show a wireframe rendering of the CME
superposed on the observed image, and the bottom panels are
simulated white-light images.}
\end{figure}

\clearpage

\begin{figure}
\epsscale{1} \plotone{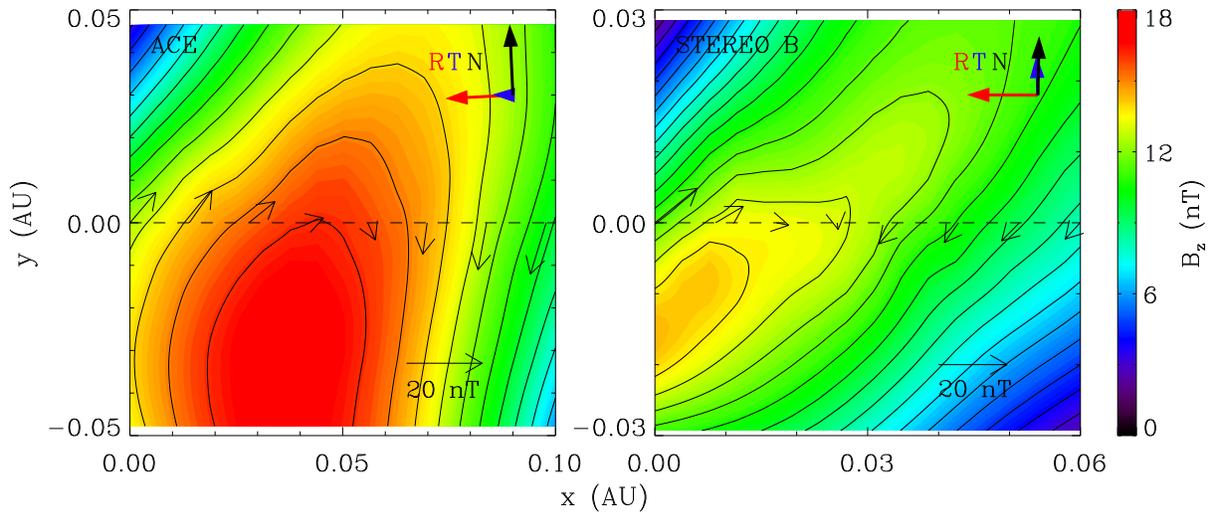} \caption{Reconstructed cross sections
of the MC at ACE (left) and STEREO B (right). Black contours show
the distribution of the vector potential, and the color shading
indicates the value of the axial magnetic field. The dashed line
marks the trajectory of each spacecraft (which can be used as a
proxy for the ecliptic plane projected on the cross section). The
thin black arrows denote the direction and magnitude of the observed
magnetic fields projected onto the cross section, and the thick
colored arrows show the projected RTN directions.}
\end{figure}

\clearpage

\begin{figure}
\epsscale{0.7} \plotone{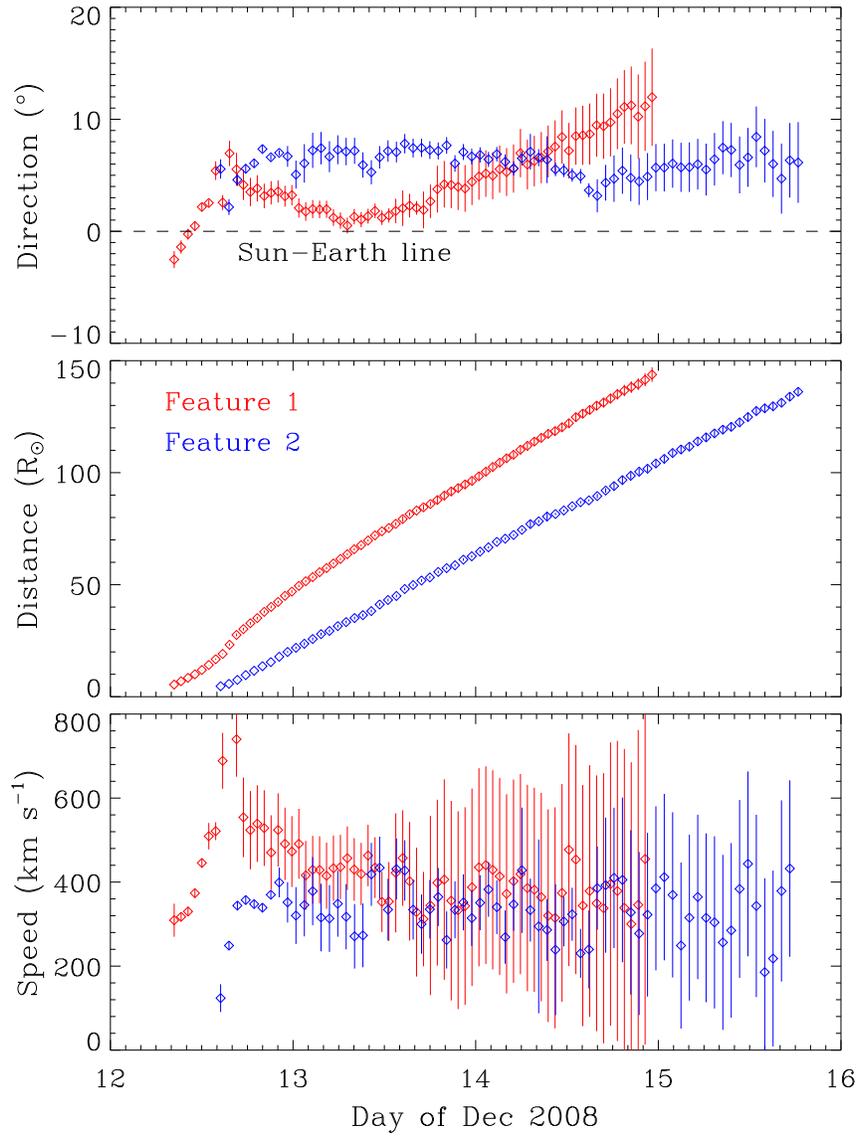} \caption{Propagation angle, radial
distance and speed of the leading (red) and trailing (blue) edges of
the 2008 December 12 CME derived from geometric triangulation
analysis. Similar format to Figure~4.}
\end{figure}

\clearpage

\begin{figure}
\epsscale{0.8} \plotone{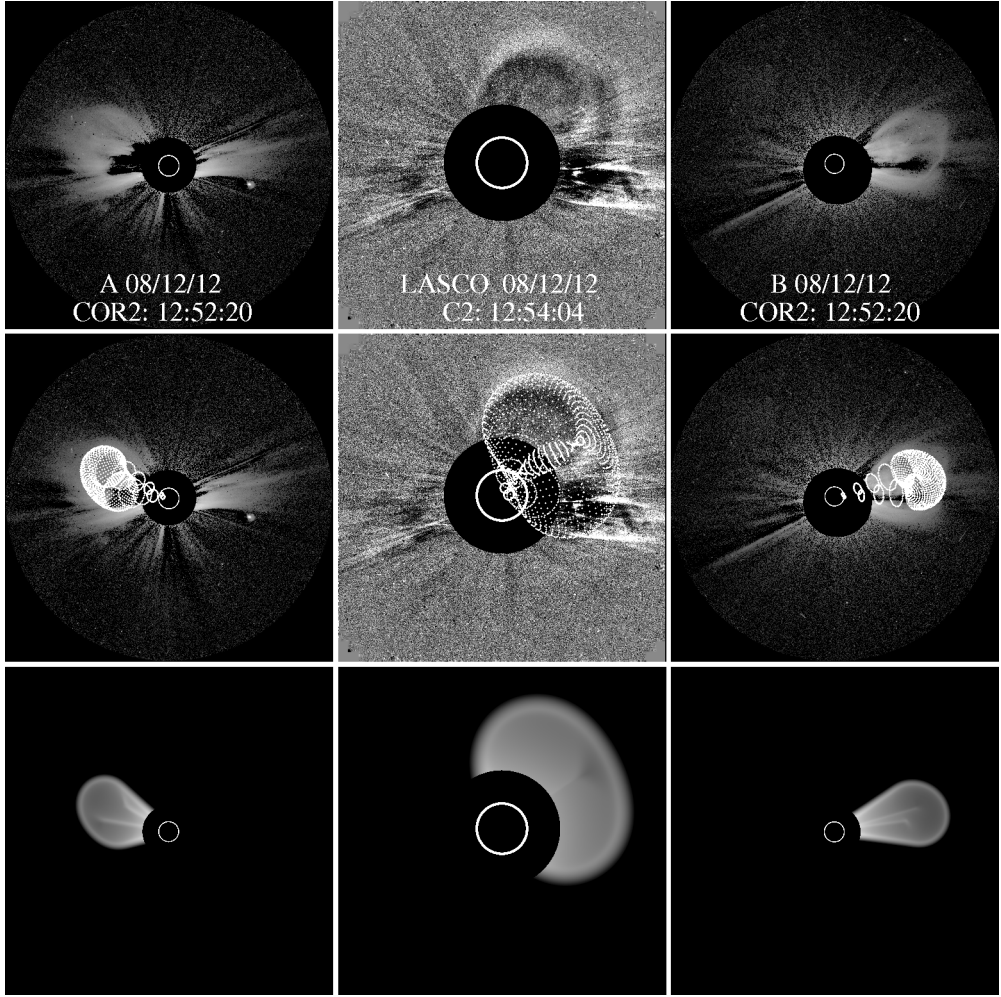} \caption{Similar format to
Figure~7, but for the 2008 December 12 CME.}
\end{figure}

\clearpage

\begin{figure}
\epsscale{0.6} \plotone{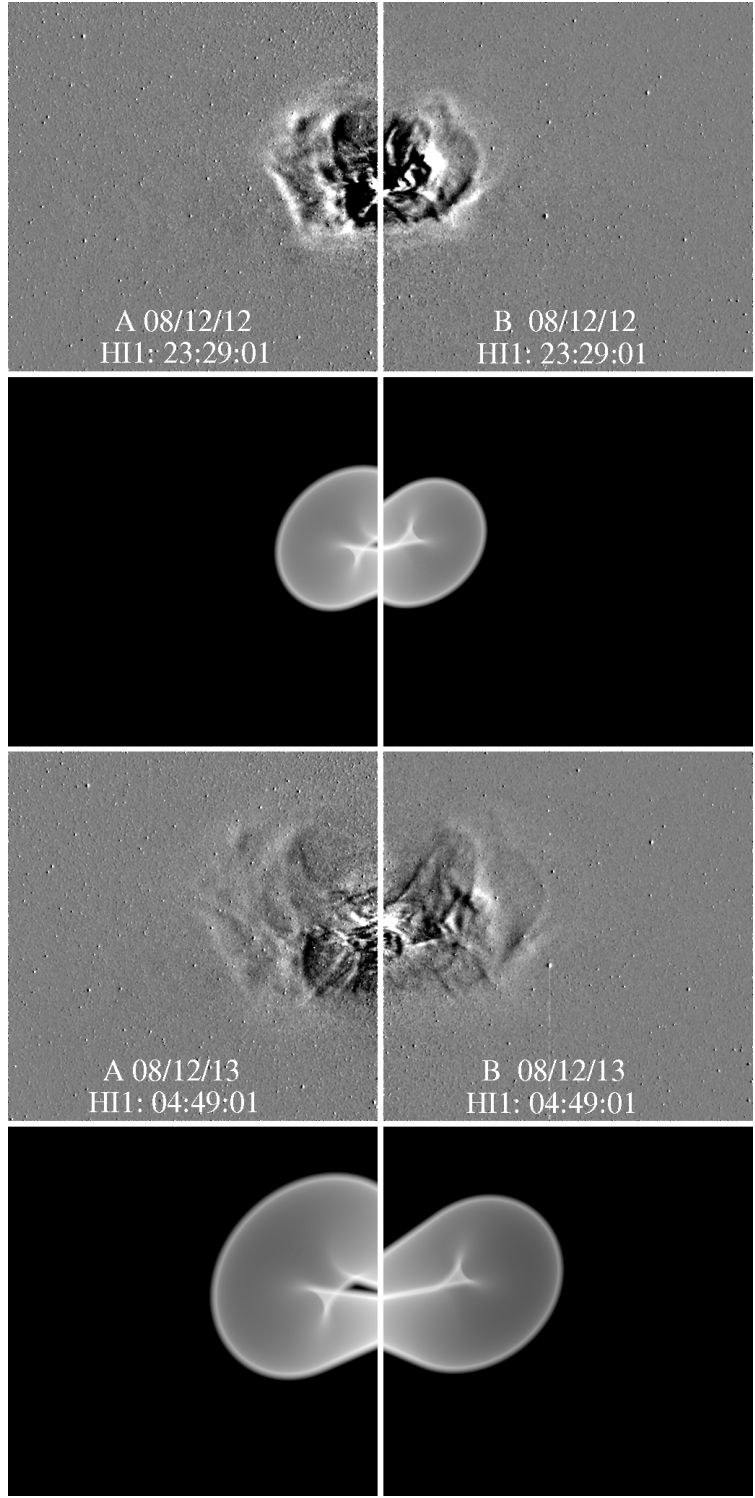} \caption{Comparison between HI1
observations and corresponding simulations for the 2008 December 12
CME. Views from STEREO A (left) and B (right) at two times are shown
by the top and bottom panels, respectively.}
\end{figure}

\clearpage

\begin{figure}
\epsscale{0.7} \plotone{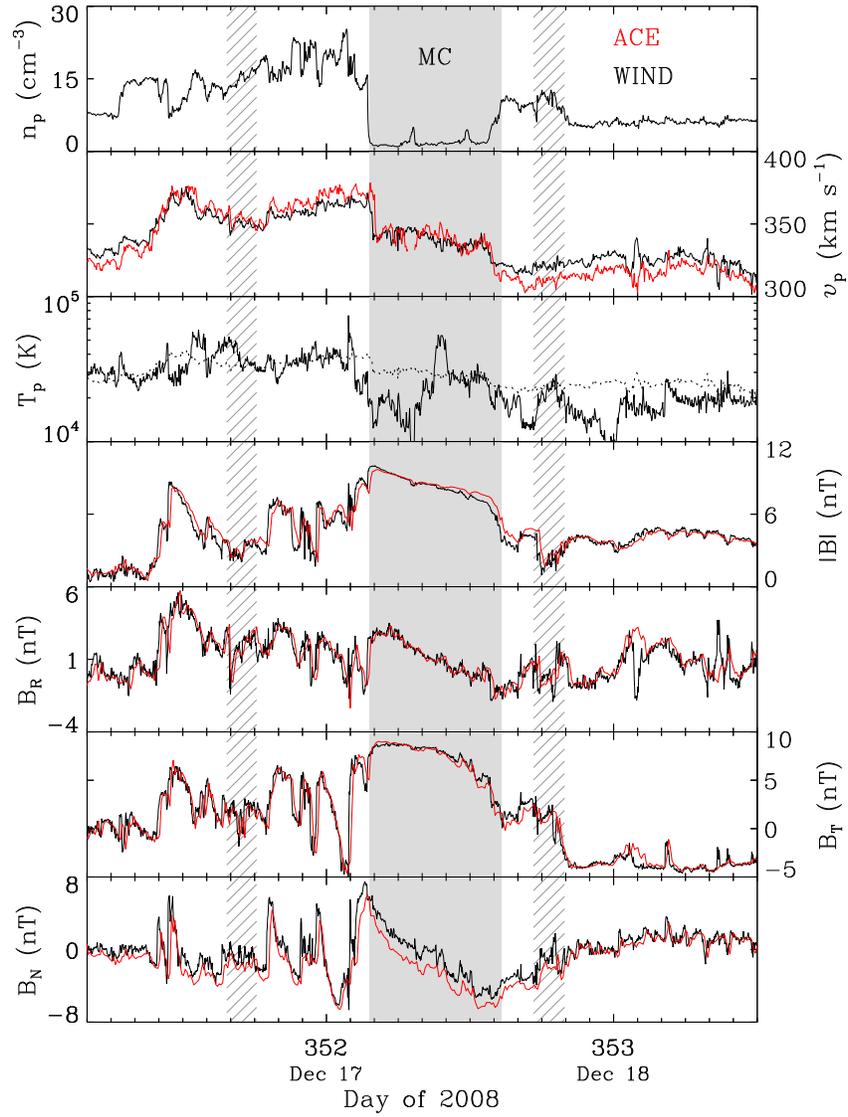} \caption{Similar format to
Figure~5, but for the measurements of the 2008 December 17 MC at
WIND (black) and ACE (red). The hatched area shows the arrival times
(with uncertainties) of the CME leading and trailing edges predicted
by the geometric triangulation technique.}
\end{figure}

\clearpage

\begin{figure}
\epsscale{0.8} \plotone{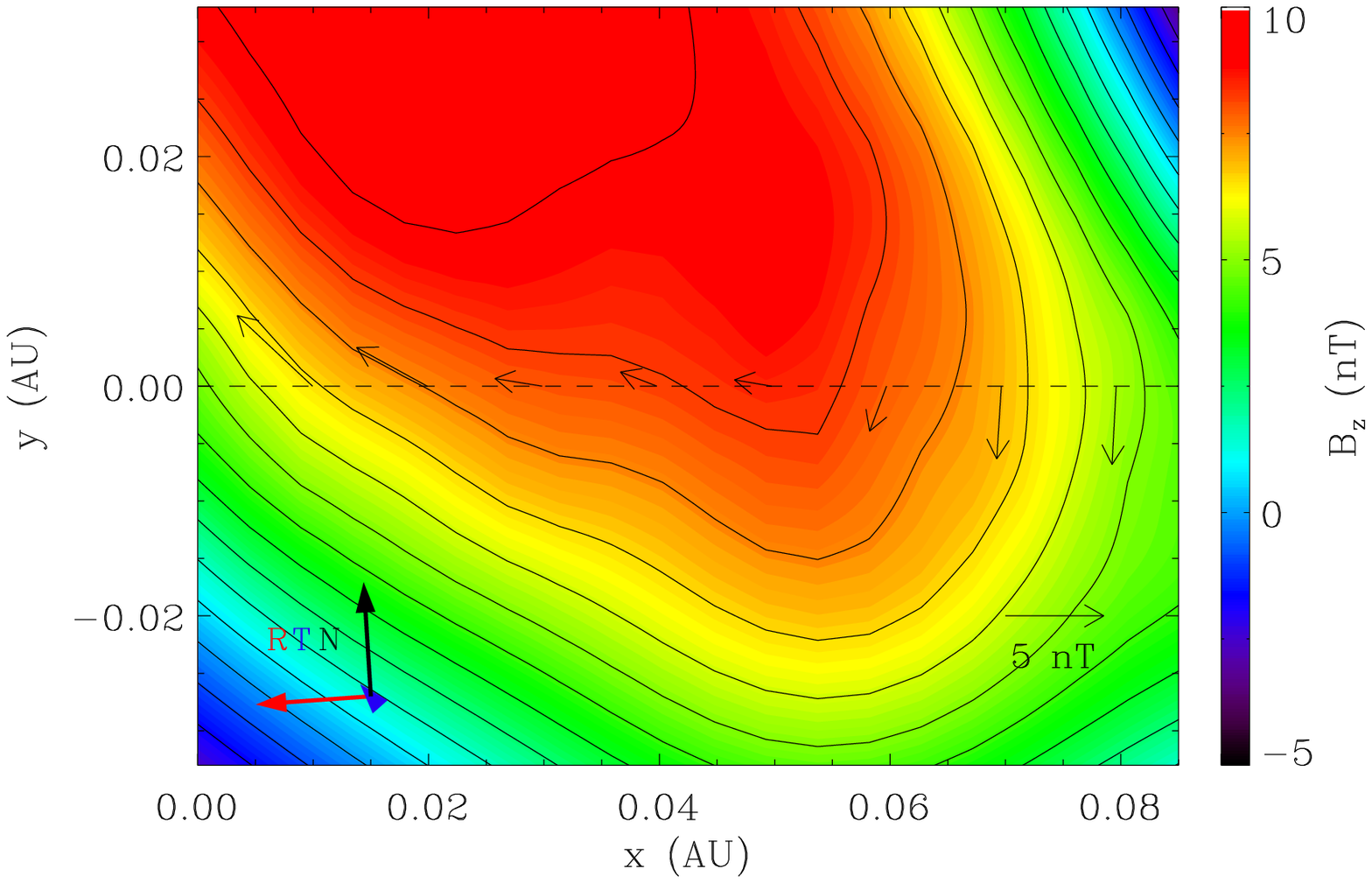} \caption{Cross section of the 2008
December 17 MC reconstructed from WIND measurements. Similar format
to Figure~8.}
\end{figure}

\clearpage

\begin{figure}
\epsscale{0.8} \plotone{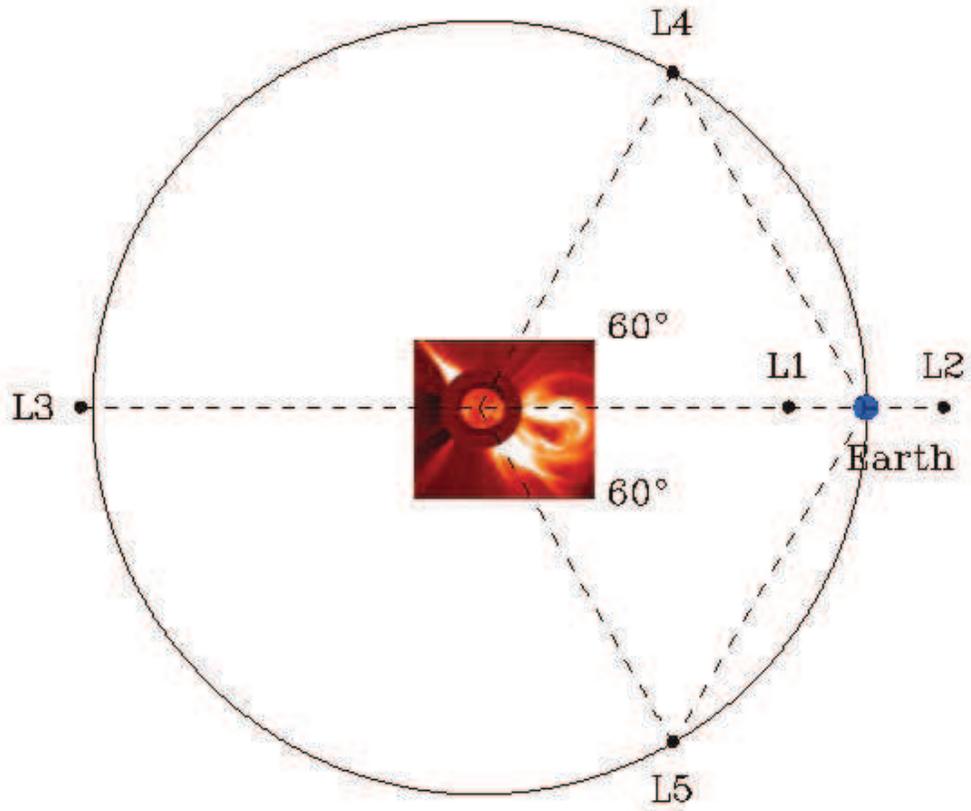} \caption{The five Lagrangian points
of the Sun-Earth system in the ecliptic plane (not to scale). The
circle represents the orbit of the Earth. L4 and L5, which form the
apex of two equilateral triangles, are well situated for geometric
triangulation of CME observations.}
\end{figure}

\clearpage

\begin{figure}
\epsscale{1} \plotone{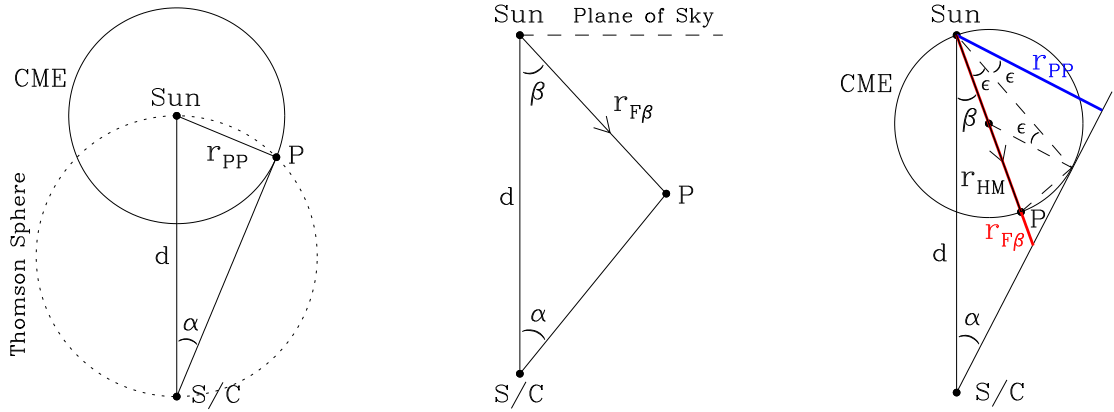} \caption{CME geometry assumed to
convert elongation angles to radial distances. Left: Point P
approximation assuming a spherical front centered at the Sun.
Middle: Fixed $\beta$ approximation assuming a compact structure
moving along a fixed direction. Right: Harmonic mean approximation
assuming a spherical front attached to the Sun moving along a
constant direction.}
\end{figure}

\clearpage

\begin{figure}
\epsscale{0.8} \plotone{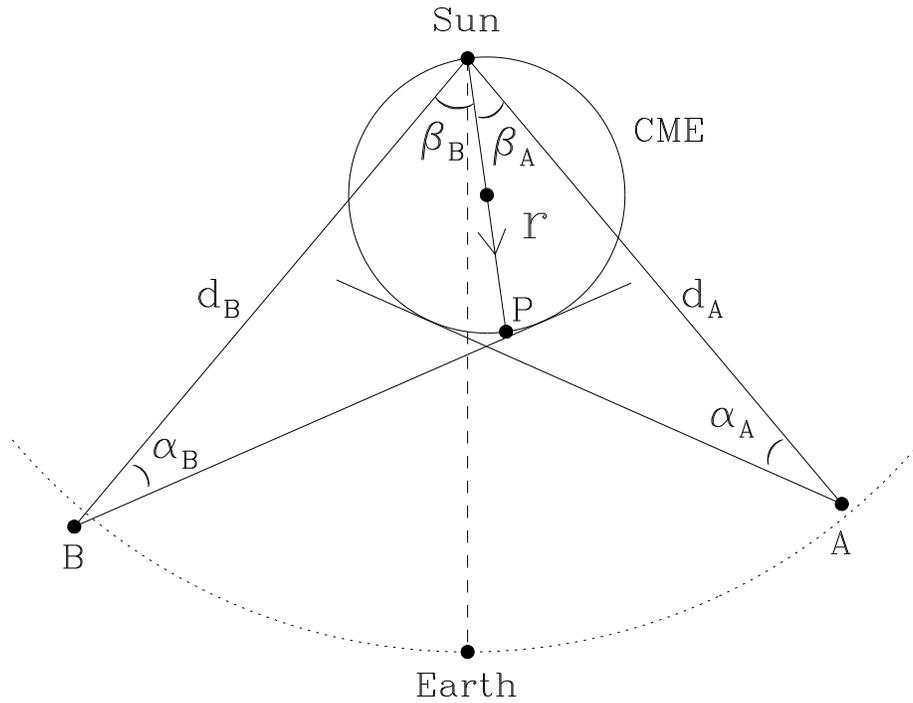} \caption{Similar format to
Figure~1, but for the geometric triangulation under the harmonic
mean approximation. A spherical front attached to the Sun is assumed
for the CME geometry. Only the case that the CME nose is propagating
between the two spacecraft is shown. Note that, under the harmonic
mean approximation, CMEs with nose outside the two spacecraft may
also be seen by the HIs on both spacecraft.}
\end{figure}

\end{document}